\documentclass[10pt,conference]{IEEEtran}

\usepackage{amsmath, amsfonts, amssymb}
\usepackage{tikz}
\usepackage{caption}
\usepackage{subcaption}
\usepackage{graphicx}
\usepackage{mathrsfs} 

\usepackage{amsthm}

\usepackage{hyperref}
\usepackage{cleveref}

\newtheoremstyle{customdefinition}
  {3pt} 
  {3pt} 
  {\itshape} 
  {} 
  {\bfseries} 
  {:} 
  {.5em} 
  {} 

\theoremstyle{customdefinition}
\newtheorem{definition}{Definition}

\newtheoremstyle{customexample}
  {3pt} 
  {3pt} 
  {\itshape} 
  {} 
  {\bfseries} 
  {:} 
  {.5em} 
  {} 

\theoremstyle{customexample}
\newtheorem{example}{Example}

\newcommand{\mydefinition}[3]{
  \begin{definition}
    \label{#1}
    \textbf{#2.} #3
  \end{definition}
}

\newcommand{\myexample}[3]{
  \begin{example}
    \label{#1}
    \textbf{#2.} #3
  \end{example}
}

\newcommand{\scalar}[1]{#1} 
\newcommand{\qubit}[1]
{\mathbf{#1}} 
\newcommand{\graph}[1]{\mathcal{#1}} 
\newcommand{\matr}[1]{\mathfrak{#1}} 

\newcommand{\set}[1]{\mathbb{#1}} 

\newcommand{\neigbors}[1]{\set {N}_\textbf{#1}}

\newcommand{\function}[1]{\mathit{#1}} 
\newcommand{\dict}[1]{\boldsymbol{#1}} 

\newcommand{\rootqubit}[1]{\qubit{q^{\text{r}}_{#1}}}
\newcommand{\terminalqubit}[1]{\qubit{q^{\text{t}}_{#1}}}

\newcommand{\lqubit}[1]{\qubit{q_{#1}}}
\newcommand{\vqubit}[1]{\qubit{z_{#1}}}

\usepackage{algorithm}
\usepackage{algpseudocode}
\usepackage{graphicx} 
\usepackage{amsmath}
\usepackage{xcolor}
\usepackage{tikz}
\usepackage{subcaption}

\usepackage{mdframed}

\usepackage{booktabs} 
\usepackage{caption} 

\usepackage{hyperref}
\usepackage{cite}

\usepackage[braket, qm]{qcircuit}
\usepackage{graphicx}

\usetikzlibrary{matrix,calc}

\definecolor{lightblue}{RGB}{173,216,230}

\usepackage{xcolor}  

\definecolor{bluegreen}{HTML}{009E73}
\definecolor{skyblue}{HTML}{56B4E9}
\definecolor{paleviolet}{HTML}{CC79A7}

\usepackage{algorithm}
\usepackage{algpseudocode}
\usepackage{graphicx} 
\usepackage{amsmath}
\usepackage{xcolor}
\usepackage{tikz}

\usetikzlibrary{matrix,calc}
\usepackage{comment}

\usepackage{eso-pic}

\usepackage{pict2e}

\usepackage{color}

\makeatletter
\newcommand*\node[2]{%
  \protected@edef\@currentlabel{#2}%
  \label{#1}%
  #2%
}
\makeatother

\usepackage{tikz}
\usetikzlibrary{tikzmark}
\usetikzlibrary{tikzmark,calc}

\usepackage{tikz}
\usetikzlibrary{calc}
\usetikzlibrary{positioning}

\usepackage[braket, qm]{qcircuit}

\usepackage[utf8]{inputenc}
\usepackage{amsmath}
\usepackage{listings}
\usepackage{color}
\usepackage{graphicx}
\usepackage{hyperref}
\usetikzlibrary{shapes.geometric, arrows, positioning}

\tikzstyle{startstop} = [rectangle, rounded corners, minimum width=3cm, minimum height=1cm,text centered, draw=black, fill=red!30]
\tikzstyle{io} = [trapezium, trapezium left angle=70, trapezium right angle=110, minimum width=3cm, minimum height=1cm, text centered, draw=black, fill=blue!30]
\tikzstyle{process} = [rectangle, minimum width=3cm, minimum height=1cm, text centered, draw=black, fill=orange!30]
\tikzstyle{decision} = [diamond, aspect=2, minimum width=3cm, minimum height=1cm, text centered, draw=black, fill=green!30]
\tikzstyle{arrow} = [thick,->,>=stealth]

\definecolor{codegreen}{rgb}{0,0.6,0}
\definecolor{codegray}{rgb}{0.5,0.5,0.5}
\definecolor{codepurple}{rgb}{0.58,0,0.82}
\definecolor{backcolour}{rgb}{0.95,0.95,0.92}

\lstdefinestyle{mystyle}{
    backgroundcolor=\color{backcolour},   
    commentstyle=\color{codegreen},
    keywordstyle=\color{magenta},
    numberstyle=\tiny\color{codegray},
    stringstyle=\color{codepurple},
    basicstyle=\ttfamily\footnotesize,
    breakatwhitespace=false,         
    breaklines=true,                 
    captionpos=b,                    
    keepspaces=true,                 
    numbers=left,                    
    numbersep=5pt,                  
    showspaces=false,                
    showstringspaces=false,
    showtabs=false,                  
    tabsize=2
}

\title{Graph-based identification of qubit network (GidNET) for qubit reuse}

\author{\IEEEauthorblockN{Gideon Uchehara, Tor M. Aamodt, Olivia Di Matteo}
\IEEEauthorblockA{\textit{Electrical and Computer Engineering} \\
\textit{The University of British Columbia}\\
Vancouver, Canada \\
\{ gideon.uchehara, aamodt, olivia \}@ece.ubc.ca}
}

\date{\today}

\begin{document}

\maketitle

\begin{abstract}
Quantum computing introduces the challenge of optimizing quantum resources
crucial for executing algorithms within the limited qubit availability of
current quantum architectures. Existing qubit reuse algorithms face a trade-off
between optimality and scalability, with some achieving optimal reuse but
limited scalability due to computational complexities, while others exhibit
reduced runtime at the expense of optimality.  This paper introduces GidNET
(Graph-based Identification of qubit NETwork), an algorithm for optimizing qubit
reuse in quantum circuits. By analyzing the circuit's Directed Acyclic Graph
(DAG) representation and its corresponding candidate matrix, GidNET identifies
higher-quality pathways for qubit reuse more efficiently. Through a comparative
study with established algorithms, notably QNET ~\cite{fang2023dynamic}, GidNET
not only achieves a consistent reduction in compiled circuit widths by a geometric mean of 4.4\%, reaching up to 21\% in larger
circuits, but also demonstrates enhanced computational speed and scaling, with
average execution time reduction of 97.4\% (i.e., 38.5$\times$ geometric mean speedup) and up to 99.3\% (142.9$\times$ speedup) across various circuit sizes. Furthermore, GidNET consistently outperforms Qiskit in circuit width reduction, achieving an average improvement of 59.3\%, with maximum reductions of up to 72\% in the largest tested circuits. These results demonstrate GidNET's ability to improve circuit width and runtime, offering a solution for quantum computers with limited numbers of qubits.
\end{abstract}

\section{Introduction}
Central to quantum computing is the efficient execution of quantum circuits, a fundamental abstraction underpinning the expression of many quantum algorithms. This necessitates quantum circuit compilation, wherein high-level quantum algorithms are transformed into optimized instructions tailored to a target quantum processor. Essential factors in this optimization include the number of qubits, their connectivity, gate fidelity, and error rates, all aimed at minimizing computational resources and errors during algorithm execution.

At the heart of circuit optimization lies the strategic management of qubits, particularly through their reuse. The innovation of qubit reuse algorithms, which re-purpose qubits in a circuit to a dynamic format requiring fewer resources, marks a significant stride in quantum computing ~\cite{DCKFF22, fang2023dynamic, sadeghi2022quantum}. This approach, bolstered by the advent of mid-circuit measurement and reset, enhances the scalability and efficiency of computations across both trapped-ion~\cite{pino2021demonstration} and superconducting systems~\cite{IBMQuantumDynamicCircuits, corcoles2021exploiting}. Transitioning from static to dynamic quantum circuits, which adjust in real-time to measurement outcomes, qubit reuse is essential across various applications, from error correction~\cite{ryan2022implementing} and tensor network state preparation~\cite{chertkov2022holographic} to entanglement spectroscopy~\cite{yirka2021qubit} and quantum machine learning~\cite{harvey2023sequence}. By mitigating quantum resource demands, it plays a crucial role in overcoming scalability challenges, potentially enhancing circuit fidelity ~\cite{BPK23} and reducing complexities ~\cite{HJC+23}. The industry's movement towards dynamic circuit capabilities, evidenced by recent hardware enhancements~\cite{IBMQuantumDynamicCircuits, pino2021demonstration}, underscores the role of qubit reuse in advancing quantum computing.

The quest for efficient qubit reuse has led to the development of two primary algorithmic approaches: exact algorithms and greedy heuristics ~\cite{DCKFF22}. Exact algorithms, utilizing the Constraint Programming and Satisfiability (CP-SAT) model, offer precision and serve as a valuable benchmark for small-scale systems, albeit with limited scalability. Conversely, greedy heuristic algorithms demonstrate polynomial growth in qubit number and rapid execution times, at the expense of not always achieving optimal results. 

In this paper we propose the graph-based identification of qubit network
(GidNET) algorithm, a technique to bridge the gap between the exactitude of
CP-SAT models and the scalability of greedy heuristics. GidNET leverages the
Directed Acyclic Graph (DAG) of a quantum circuit and its matrix representation
to identify pathways for qubit reuse. We demonstrate through
simulation that, compared to the QNET algorithm in ~\cite{fang2023dynamic} and the Qiskit implementation of the qubit reuse algorithm in ~\cite{DCKFF22}, which we henceforth refer to as Qiskit, GidNET not only achieves a consistent reduction in compiled circuit width but also demonstrates enhanced computational speed.
Our contributions are summarized as follows:

\begin{itemize}
    \item Introduction of `\emph{reuse sequences}' determined through common neighbors in the circuit's DAG leading to higher-quality pathways for qubit reuse more efficiently.
    \item Achievement of consistently reduced compiled circuit widths by a geometric mean of 4.4\% and up to 21\% in larger circuits, surpassing existing algorithms ~\cite{DCKFF22, fang2023dynamic}.
    \item Significant acceleration of computational runtime over \cite{fang2023dynamic}, with a geometric
mean improvement of 97.4\% (or 38.5$\times$), reaching up to 99.3\% (142.9$\times$) across different circuit sizes.
    \item Validation of GidNET's theoretical complexity through polynomial regression and F-tests, confirming its efficacy against empirical data.
\end{itemize}

By optimizing qubit reuse, GidNET not only extends the capabilities of existing
quantum systems but also reduces operational overhead and is
expected to lead to improved quality of circuit execution.

This paper is structured as follows. In Section~\ref{sec:related_work}, we review existing qubit reuse algorithms and introduce GidNET's approach. In Section~\ref{sec:data_structures}, we detail the data structures employed by GidNET to facilitate efficient reuse. Section~\ref{sec:algorithm} describes the algorithm's implementation and computational complexity. Section~\ref{sec:experimental_results} presents the empirical validation of GidNET, demonstrating performance improvements over existing algorithms. Finally, Section~\ref{sec:conclusion} summarizes our findings and discusses potential future enhancements to GidNET. All the code referenced in this paper is publicly accessible on GitHub\footnote{https://github.com/QSAR-UBC/GidNET-qubit-reuse-algorithm}

\section{Related Work and the Novelty of GidNET} \label{sec:related_work}
In this section, we present quantum circuit cutting and explore existing algorithms for qubit reuse and their limitations, and highlight the unique features of GidNET. We discuss how GidNET leverages graph properties of a quantum circuit to determine reuse, setting it apart from other methods in terms of scalability and performance.

To maximize the utility of current noisy intermediate-scale quantum (NISQ) devices, techniques such as quantum circuit cutting and qubit reuse were developed. Algorithms for circuit cutting split the original quantum circuit either by cutting wires ~\cite{peng, uchehara2022rotation, Perlin2021, tang, Lowe2022, brenner2023optimal, harada2023optimal}, gates ~\cite{piveteau2022circuit, mitarai2021overhead, ufrecht2023cutting, mitarai2021constructing}, or both ~\cite{brandhofer2023optimalcutting} into smaller subcircuits that can be run separately, either on the same quantum device or on different devices\cite{DiMatteo2021}. The results are post-processed and combined using a classical computer. This approach has several drawbacks. For example, the runtime for circuit cutting increases exponentially in the number of cuts on a circuit. In contrast, qubit reuse leverages the ability of a quantum computer to perform mid-circuit measurements and resets, allowing qubit to be reused after they are reset, without necessitating exponential overhead~\cite{DCKFF22}.

Recent efforts have focused on algorithms and techniques to enhance qubit reuse and overall quantum circuit efficiency ~\cite{DCKFF22, fang2023dynamic, sadeghi2022quantum, hua2023caqr, pawar2023integrated, niu2023powerful,brandhofer2023optimal}. Qubit reuse is particularly appealing because promising algorithms for integer factorization ~\cite{shor1994algorithms}, database search ~\cite{grover1996fast}, machine learning ~\cite{biamonte2017quantum}, and other applications require a large number of qubits. This demand exceeds the capacities of current NISQ computers. The concept of qubit reuse, specifically through wire recycling, was pioneered by Paler et al. \cite{PWD16}, emphasizing the underutilization of qubits across quantum circuit computations. This seminal work laid the foundation for subsequent research in qubit reuse strategies. DeCross et al. \cite{DCKFF22} advanced the field by addressing qubit reuse through mid-circuit measurements and resets, presenting both an exact optimization model and a heuristic for large-scale quantum circuits. Their work highlighted the dual circuit concept, underscoring the symmetry in qubit requirements between a circuit and its temporal reverse. Recent developments by Hua et al. \cite{HJC+23} introduced a compiler-assisted tool leveraging dynamic circuit capabilities for qubit reuse. Their empirical analysis demonstrated notable improvements in qubit efficiency and circuit fidelity on actual quantum hardware. Brandhofer et al. \cite{BPK23} proposed an SAT-based model for qubit reuse optimization, tackling the computational challenges inherent in scaling qubit reuse methods. Their approach aims to achieve optimal circuit configurations with respect to various performance metrics, including circuit depth and swap gate insertions. 

Most recently, Fang et al. \cite{fang2023dynamic} presented a comprehensive study on dynamic quantum circuit compilation, offering the first general framework for optimizing this process through graph manipulation. They formulated the problem using binary integer programming, and introduced algorithms for assessing circuit reducibility and devising dynamic compilation schemes. Their comparative analysis highlights the superior performance of their methods, providing a rigorous foundation for future advances in dynamic circuit compilation ~\cite{fang2023dynamic}. 

Building on these works, but particularly inspired by both the complexity of
current quantum challenges and the innovative approach by Fang et
al. ~\cite{fang2023dynamic}, we propose GidNET, a qubit reuse algorithm that
leverages advanced analysis techniques of the quantum circuit's graph structure,
which many of the related works above, apart from QNET~\cite{fang2023dynamic},
do not exploit. Specifically, GidNET utilizes the circuit DAG and the candidate
matrix derived from its biadjacency graph. It uniquely determines a
greedily-optimized reuse sequence for each qubit, leveraging a property we call
common neighbors to methodically select a sequence of qubits for reuse based on interactions between qubits. GidNET ensures that each decision on qubit reuse is informed by a clear understanding of the existing network of qubit relationships, streamlining the application of reuse and improving the outcome of circuit compilation. This strategic approach contrasts with ~\cite{fang2023dynamic}, which broadly evaluates every edge in the candidate matrix to identify the edge that maximizes the possibility of adding more edges in subsequent steps in each iteration.

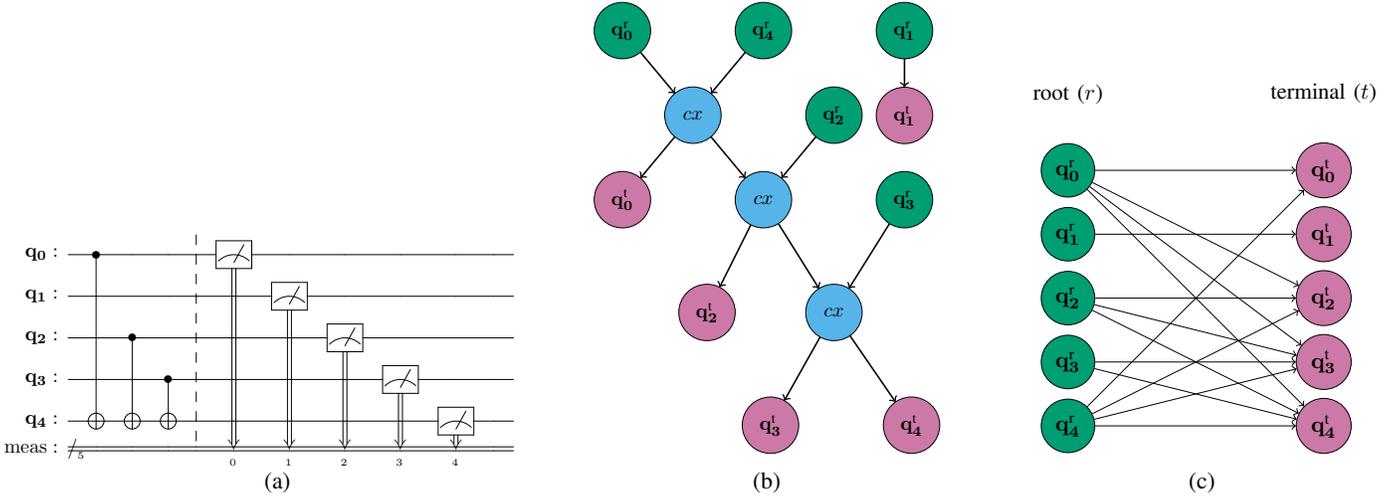
\begin{figure*}[htbp]
    \centering
    \hspace*{1.0mm} 
    \begin{subfigure}{0.35\textwidth}
        \centering
        \scalebox{0.75}{
        \Qcircuit @C=1.0em @R=0.7em {
        & \lstick{\lqubit 0 :  } & \ctrl{4} & \qw & \qw \barrier[0em]{4} & \qw & \meter & \qw & \qw & \qw & \qw & \qw & \qw \\
        & \lstick{\lqubit 1 :  } & \qw & \qw & \qw & \qw & \qw & \meter & \qw & \qw & \qw & \qw & \qw\\
        & \lstick{\lqubit 2 :  } & \qw & \ctrl{2} & \qw & \qw & \qw & \qw & \meter & \qw & \qw & \qw & \qw\\
        & \lstick{\lqubit 3 :  } & \qw & \qw & \ctrl{1} & \qw & \qw & \qw & \qw & \meter & \qw & \qw & \qw\\
        & \lstick{\lqubit 4 :  } & \targ & \targ & \targ & \qw & \qw & \qw & \qw & \qw & \meter & \qw & \qw\\
        & \lstick{\mathrm{meas} :  } & \lstick{/_{_{5}}} \cw & \cw & \cw & \cw & \dstick{_{_{\hspace{0.0em}0}}} \cw \ar @{<=} [-5,0] & \dstick{_{_{\hspace{0.0em}1}}} \cw \ar @{<=} [-4,0] & \dstick{_{_{\hspace{0.0em}2}}} \cw \ar @{<=} [-3,0] & \dstick{_{_{\hspace{0.0em}3}}} \cw \ar @{<=} [-2,0] & \dstick{_{_{\hspace{0.0em}4}}} \cw \ar @{<=} [-1,0] & \cw & \cw\\
        }
        }
        \caption{}
        \label{fig:quantum_circuit}
    \end{subfigure}%
    \hfill
    \begin{subfigure}{0.33\textwidth}
        \centering
        \scalebox{0.75}{
        \begin{tikzpicture}
          [node distance=1.5cm, every join/.style={->, thick, shorten >=1pt},
          input/.style={circle, draw=black, fill=bluegreen, minimum size=10mm, outer sep=0pt},
          output/.style={circle, draw=black, fill=paleviolet, minimum size=10mm, outer sep=0pt},
          cnot/.style={circle, draw=black, fill=skyblue, minimum size=10mm, outer sep=0pt}]
        
          \node [input] (q0) {\(\rootqubit 0\)};
          \node [input, right=of q0] (q4) {\(\rootqubit 4\)};
          \node [input, right=of q4] (q1) {\(\rootqubit 1\)};
          \node [cnot] (c1) at ($(q0)!0.5!(q4) + (0,-1.5)$) {\(cx\)};
          \node [output] (q0out) at ($(0,-3.0)$) {\(\terminalqubit 0\)};
          \node [input, right=of c1] (q2) {\(\rootqubit 2\)};
          \node [output, below of=q1, yshift=-0.005cm] (q1out) {\(\terminalqubit 1\)};
          \node [cnot] (c2) at ($(c1)!0.5!(q2) + (0,-1.5)$) {\(cx\)};
          \node [output] (q2out) at ($(1.5,-5.0)$) {\(\terminalqubit 2\)};
          \node [input, right=of c2] (q3) {\(\rootqubit 3\)};
          \node [cnot] (c3) at ($(c2)!0.5!(q3) + (0,-2.0)$) {\(cx\)};
          \node [output] (q3out) at ($(q2out)!0.5!(c3) + (0,-2.0)$) {\(\terminalqubit 3\)};
          \node [output, right=of q3out] (q4out) {\(\terminalqubit 4\)};
          
          \draw[->, thick] (q0) -- (c1);
          \draw[->, thick] (q4) -- (c1);
          \draw[->, thick] (q1) -- (q1out);
          \draw[->, thick] (c1) -- (q0out);
          \draw[->, thick] (c1) -- (c2);
          \draw[->, thick] (q2) -- (c2);
          \draw[->, thick] (c2) -- (q2out);
          \draw[->, thick] (c2) -- (c3);
          \draw[->, thick] (q3) -- (c3);
          \draw[->, thick] (c3) -- (q3out);
          \draw[->, thick] (c3) -- (q4out);
        \end{tikzpicture}
        }
        \caption{}
        \label{fig:bv_cct_dag}
    \end{subfigure}%
    \hfill
    \begin{subfigure}{0.27\textwidth}
        \centering
        \scalebox{0.85}{
        \begin{tikzpicture}
          [every join/.style={->, thick, shorten >=1pt},
        node distance=2cm, 
        main/.style = {draw, circle}] 
        \foreach \x in {0,...,4}
        {
            \node[main, fill=bluegreen] (s\x) at (0,-\x) {\(\rootqubit \x\)};
            \node[main, fill=paleviolet] (t\x) at (4,-\x) {\(\terminalqubit \x\)};
        }
        
        \draw[->] (s0) -- (t0);
        \draw[->] (s0) -- (t2);
        \draw[->] (s0) -- (t3);
        \draw[->] (s0) -- (t4);
        
        \draw[->] (s1) -- (t1);
        \draw[->] (s2) -- (t2);
        \draw[->] (s2) -- (t3);
        \draw[->] (s2) -- (t4);
        
        \draw[->] (s3) -- (t3);
        \draw[->] (s3) -- (t4);
        
        \draw[->] (s4) -- (t0);
        \draw[->] (s4) -- (t2);
        \draw[->] (s4) -- (t3);
        \draw[->] (s4) -- (t4);
        
        \node[above=0.5cm of s0] (Source) {root (\(r\))};
        \node[above=0.5cm of t0] (Target) {terminal (\(t\))};
        \end{tikzpicture}
        }
        \caption{}
        \label{fig:biadjacency_graph}
    \end{subfigure}
    \caption{(a) A 5-qubit quantum circuit. (b) DAG, \(\graph
G\) for the circuit in ~\autoref{fig:quantum_circuit}. Green nodes are the roots
of qubits while violet nodes are terminals of qubits. Blue nodes are CNOT gates
between two qubits. (c) Simplified DAG (biadjacency graph) \(\graph G'\),
derived from ~\autoref{fig:bv_cct_dag} showing connections from roots (\(r\)) to
terminals (\(t\)) of qubits. An edge, (\(\rootqubit i, \terminalqubit j\))
between two qubits indicates a direct path from qubit \(\lqubit {{i}}\) to qubit
\(\lqubit {{j}}\) in \(\graph G\), highlighting qubit pairs ineligible for reuse.}
    \label{fig:multi_panel}
\end{figure*}

\section{Data Structures} \label{sec:data_structures}
This section summarizes the framework for circuit analysis and qubit reuse introduced by ~\cite{fang2023dynamic} that is leveraged by GidNET. Given a circuit \(\graph Q\), with set of \(\scalar{n}\) qubits \(\{\lqubit {{0}}, ...,  \lqubit {{n-1}}\}\), the first step is to compute the circuit's DAG, \( \graph{G=(V, E)}\), which encapsulates the causal sequence of quantum operations. An example circuit and its DAG are shown in \autoref{fig:quantum_circuit} and \autoref{fig:bv_cct_dag} respectively. The DAG consists of three types of vertices, \( \graph V \): roots (green), terminals (violet), and operations (blue). The roots are vertices with no incoming edges, representing the first layer of quantum operations on each qubit, usually reset operations. The terminals are vertices with no outgoing edges, representing the last layer of quantum operations, typically quantum measurements. Internal vertices with incoming and outgoing edges correspond to intermediate quantum operations, such as quantum gates. Directed edges between these vertices represent qubits on which these operations are applied.

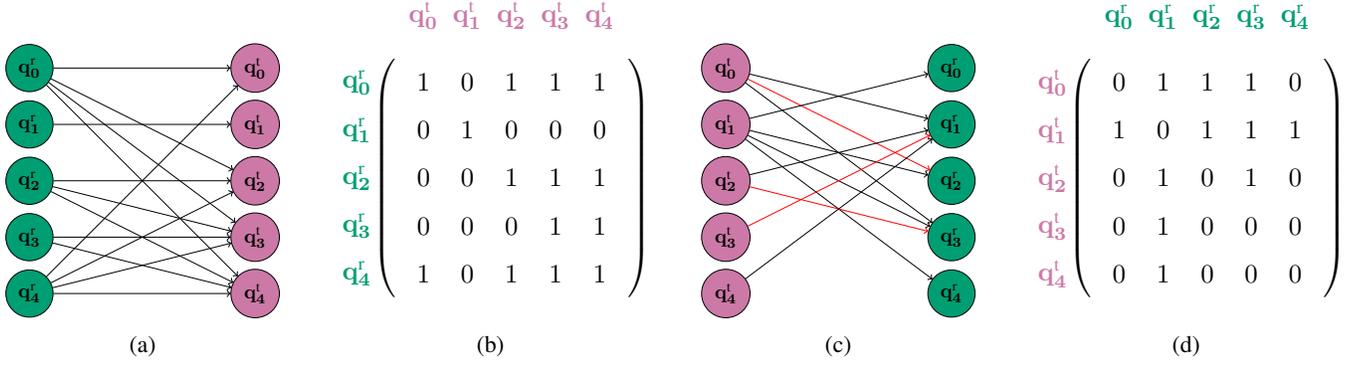
\begin{figure*}[htbp]
    \centering
    \begin{subfigure}{0.24\textwidth}
        \centering
        \scalebox{0.75}{
        \begin{tikzpicture}
            [every join/.style={->, thick, shorten >=1pt},
            node distance=1.5cm, 
            main/.style = {draw, circle}] 
            \foreach \x in {0,...,4}
            {
                \node[main, fill=bluegreen] (s\x) at (0,-\x) {\(\rootqubit \x\)};
                \node[main, fill=paleviolet] (t\x) at (4,-\x) {\(\terminalqubit \x\)};
            }
            \draw[->] (s0) -- (t0);
            \draw[->] (s0) -- (t2);
            \draw[->] (s0) -- (t3);
            \draw[->] (s0) -- (t4);
            \draw[->] (s1) -- (t1);
            \draw[->] (s2) -- (t2);
            \draw[->] (s2) -- (t3);
            \draw[->] (s2) -- (t4);
            \draw[->] (s3) -- (t3);
            \draw[->] (s3) -- (t4);
            \draw[->] (s4) -- (t0);
            \draw[->] (s4) -- (t2);
            \draw[->] (s4) -- (t3);
            \draw[->] (s4) -- (t4);
        \end{tikzpicture}
        }
        \caption{}
        \label{fig:biadjacency_graph2}
    \end{subfigure}%
    \hfill
    \begin{subfigure}{0.23\linewidth}
        \centering
        \[
        \begin{tikzpicture}
        \matrix (m) [matrix of math nodes,left delimiter={(},right delimiter={)},
            column 1/.style={nodes={text=black}},
            column 2/.style={nodes={text=black}},
            column 3/.style={nodes={text=black}},
            column 4/.style={nodes={text=black}},
            column 5/.style={nodes={text=black}},
            row sep=0.5em, column sep=0.5em, 
            nodes in empty cells]{
            1 & 0 & 1 & 1 & 1  \\
            0 & 1 & 0 & 0 & 0  \\
            0 & 0 & 1 & 1 & 1 \\
            0 & 0 & 0 & 1 & 1 \\
            1 & 0 & 1 & 1 & 1 \\
            };
        
            \foreach \num in {1,...,5}{
              \node [left=1em of m-\num-1] {\textcolor{bluegreen}{\(\rootqubit {\pgfmathparse{\num-1}\pgfmathprintnumber{\pgfmathresult}}\)}};
            }
            
            \foreach \num in {1,...,5}{
              \node [above=1em of m-1-\num] {\textcolor{paleviolet}{\(\terminalqubit {\pgfmathparse{\num-1}\pgfmathprintnumber{\pgfmathresult}}\)}};
            }
        
        \end{tikzpicture}
         \]
        \caption{}
        \label{fig:biadjacency_matrix}
    \end{subfigure}%
    \hfill
    \begin{subfigure}{0.24\textwidth}
        \centering
        \scalebox{0.75}{
        \begin{tikzpicture}
            [every join/.style={->, thick, shorten >=1pt},
            node distance=1.5cm, 
            main/.style = {draw, circle}]  
            \foreach \x in {0,...,4}
            {
                \node[main, fill=paleviolet] (s\x) at (0,-\x) {\(\terminalqubit \x\)};
                \node[main, fill=bluegreen] (t\x) at (4,-\x) {\(\rootqubit \x\)};
            }
            \draw[->] (s0) -- (t1);
            \draw[->, red] (s0) -- (t2);
            \draw[->] (s0) -- (t3);
            \draw[->] (s1) -- (t0);
            \draw[->] (s1) -- (t2);
            \draw[->] (s1) -- (t3);
            \draw[->] (s1) -- (t4);
            \draw[->] (s2) -- (t1);
            \draw[->, red] (s2) -- (t3);
            \draw[->, red] (s3) -- (t1);
            \draw[->] (s4) -- (t1);
        \end{tikzpicture}
        }
        \caption{}
        \label{fig:candidate_graph}
    \end{subfigure}%
    \hfill
    \begin{subfigure}{0.23\linewidth}
        \centering
        \[
        \begin{tikzpicture}
        \matrix (m) [matrix of math nodes,left delimiter={(},right delimiter={)},
            column 1/.style={nodes={text=black}},
            column 2/.style={nodes={text=black}},
            column 3/.style={nodes={text=black}},
            column 4/.style={nodes={text=black}},
            column 5/.style={nodes={text=black}},
            row sep=0.5em, column sep=0.5em, 
            nodes in empty cells]{
            0 & 1 & 1 & 1 & 0  \\
            1 & 0 & 1 & 1 & 1  \\
            0 & 1 & 0 & 1 & 0 \\
            0 & 1 & 0 & 0 & 0 \\
            0 & 1 & 0 & 0 & 0 \\
            };
        
            \foreach \num in {1,...,5}{
              \node [left=1em of m-\num-1] {\textcolor{paleviolet}{\(\terminalqubit {\pgfmathparse{\num-1}\pgfmathprintnumber{\pgfmathresult}}\)}};
            }
            
            \foreach \num in {1,...,5}{
              \node [above=1em of m-1-\num] {\textcolor{bluegreen}{\(\rootqubit {\pgfmathparse{\num-1}\pgfmathprintnumber{\pgfmathresult}}\)}};
            }
        
        \end{tikzpicture}
         \]
        \caption{}
        \label{fig:candidate_matrix}
    \end{subfigure}
    \caption{(a) The biadjacency graph (same as ~\autoref{fig:biadjacency_graph}). (b) The biadjacency matrix \(\matr{B}\), labels rows (root) and columns (terminals) from \(\lqubit {{0}}\) to \(\lqubit {{n-1}}\). (c) The candidate graph is the graph compliment of  ~\autoref{fig:biadjacency_graph2}. An edge signifies that the root qubit can reuse the terminal qubit post-operation. (d) The candidate matrix \(\matr{C}\), as defined in ~\autoref{eq:candidate_matrix_eqn}, with rows (terminals) and columns (roots) labeled (\(\lqubit {{0}}, ...,  \lqubit {{n-1}}\)). A `1' indicates a potential reuse opportunity.}
    \label{fig:all_figures}
\end{figure*}

A circuit DAG can be further simplified to a bipartite graph \(\graph {G' = (R,
T, E)}\), where \(\graph R\) and \(\graph T\) denote the sets of roots and terminals from the DAG representation, respectively. An edge \((r, t) \in \graph E\) connects a root \(r \in \graph R\) to a terminal \(t \in \graph T\) if there exists a directed path from \(r\) to \(t\) in the original DAG, \(\graph G\).
Each \(\rootqubit{i}\) and \(\terminalqubit{i}\) represent the root and terminal
vertices of the \(i\)-th qubit respectively. The simplified DAG, \(\graph G'\)
is known as the \emph{biadjacency graph}
~\cite{fang2023dynamic}. ~\autoref{fig:biadjacency_graph} shows the biadjacency
graph of ~\autoref{fig:bv_cct_dag}, and ~\autoref{fig:biadjacency_matrix} its
matrix representation, \(\matr B\). Row and column indices correspond to the
root and terminal nodes of qubits. The biadjacency graph is crucial for the
reuse process: if the terminal node of qubit \(\lqubit j\) is connected to the
root node of qubit \(\lqubit i\) in \(\graph G'\), the physical qubit associated
with \(\lqubit i\) cannot be measured and subsequently reused to represent qubit
\(\lqubit j\) because \(\lqubit j\)'s operation depends on \(\lqubit i\). For instance, in \autoref{fig:biadjacency_graph}, qubit $\qubit{q_4}$ cannot reuse qubit $\qubit{q_0}$ due to the edge from root $\rootqubit{4}$ to terminal $\terminalqubit{0}$.

To determine which qubits can be reused, we construct a \emph{candidate matrix},
\(\matr{C}\), the adjacency matrix of a \emph{candidate graph}. Given a biadjacency matrix \(\matr{B}\)~\cite{fang2023dynamic},
\vspace{-0.7em}
\begin{equation}
    \matr{C} = \mathbf{1}_{n \times n} - \matr{B}^{T},
    \label{eq:candidate_matrix_eqn}
    \vspace{-0.7em}
\end{equation}
\noindent where \(\mathbf{1}_{n \times n}\) is an \(n \times n\) all-ones matrix. ~\autoref{fig:candidate_graph} and ~\autoref{fig:candidate_matrix} show the candidate graph and matrix of the circuit in \autoref{fig:quantum_circuit} respectively. The candidate graph is the graph complement of the biadjacency graph where the edges are from the terminals (rows of \(\matr{C}\)) to the roots (columns of \(\matr{C}\)). Edges from a given qubit's terminal are only connected to the roots of qubits that do not share a direct connection with the qubit's terminal in the biadjacency graph. An edge in the candidate graph from a qubit's terminal to another qubit's root indicates that the latter qubit can reuse the former qubit once the former qubit has completed its operation.

\section{GidNET Qubit Reuse Algorithm} \label{sec:algorithm}
In this section, we describe GidNET in detail. In ~\autoref{sec:
algorithm_description}, we explain each operational phase, showcasing its systematic approach to identifying and optimizing qubit reuse sequences using the candidate matrix. In ~\autoref{subsec:complexity_analysis_oz_gidnet} we analyze the complexity of GidNET.

A qubit reuse algorithm transforms a static quantum circuit, \(\graph Q\), into an equivalent dynamic circuit, \(\graph D\), using fewer qubits, as illustrated in \autoref{fig:static_and_dynamic_circuits}. To avoid any ambiguity, we define distinct categories of qubits: \emph{logical qubits} denoted as \(\qubit q\), are those in the original static quantum circuit; \emph{virtual qubits} denoted as \(\qubit z\), are qubits in the transformed dynamic circuit; and \emph{physical qubits} are actual qubits on quantum hardware. 

\subsection{Qubit reuse sequence}\label{sec: optimal_reuse_path_for_t}
To explain the intuition behind GidNET, we examine the \emph{reuse sequence} for each virtual qubit in \(\graph D\).

\mydefinition{def:reuse_path}{Reuse sequence}{
We define \(\set F_{ \textbf{i}}\), the \textbf{reuse sequence} of virtual qubit \(\vqubit {{i}}\) in a dynamic quantum circuit \(\graph D\), as the sequence of logical qubits \(\lqubit {{j}}\) from the original static circuit \(\graph Q\) that are mapped onto  \(\vqubit {{i}}\). The number of virtual qubits in \(\graph D\) is the \textbf{width} of \(\graph D\).
} 

\myexample{ex:reuse_path}{Reuse sequence illustration}{
    The reuse sequences of virtual qubits \(\vqubit {0}\) and \(\vqubit {1}\) in the dynamic circuit \(\graph D\) of ~\autoref{fig:dynamic_circuit} are
\( \set F_{\textbf{0}} = \{\lqubit {0}, \lqubit {2}, \lqubit {3}, \lqubit {1}\}\) and \( \set F_{\textbf{1}} = \{\lqubit {4} \} \). \(\graph D \) has width 2.}

The purpose of GidNET is to compute the \emph{longest} or most \emph{optimized}
reuse sequences for \( \graph D\) such that  its width is minimized.
The order of logical qubits in each \(\set F_{\textbf{i}}\) plays a pivotal role in the quality of the overall solution. To find the optimal set of reuse sequences that results in the smallest circuit width, one could explore all possible sequences that form valid reuse sequences. A valid reuse sequence is one that does not introduce a cycle in the DAG of \(\graph D\). Even though finding the optimal set of reuse sequences may be computationally cheap for a small quantum circuit, it gets prohibitively expensive as circuit size increases. 

To reduce the computational cost, GidNET randomly selects from the available
logical qubits to determine each \(\set F_{ \textbf{i}}\). This randomness
necessitates running the algorithm multiple times, but improves the likelihood of identifying more optimized qubit reuse sequences for \(\graph D\).

\subsection{Description of GidNET Algorithm}\label{sec: algorithm_description}
The GidNET Algorithm ~\autoref{algo: GidNET} employs Algorithm ~\autoref{algo: optimal_reuse_path_for_t} to determine a set of \emph{reuse sequences} for \(\graph D\). GidNET's inputs are a circuit, \(\graph Q\), the number of qubits in the circuit, $n$, and the number of iterations, $i$, to run the algorithm. Its output is a set containing the optimized reuse sequences with reduced width for the dynamic circuit \(\graph D\).

\begin{algorithm}
\caption{GidNET Qubit Reuse Algorithm}
\label{algo: GidNET}
\begin{algorithmic}[1]
    \Require \(\graph Q\), \textit{n, i}
    \Ensure List of qubit reuse sequences
    
    \State $ \matr{C} \gets$ \text{Compute the Candidate Matrix of \(\graph Q\)}
    \State $\set U \gets$ \(\{\{\lqubit 0\}, \{\lqubit 1\}, ..., \{\lqubit {n-1}\}\} \)
    \If{$\forall \terminalqubit i, \lqubit j \; (\matr{C}_{\terminalqubit i, \lqubit j} = 0)$}
        \State \textbf{return} \(\set U \)
        \Comment{Irreducible circuit}
    \EndIf
    
    \For{range($i$)}
        \State $\matr{C'} \gets \Call{Copy}{\matr{C}}$
        \State $\set U' \gets$ \{ \}
        
        \While{$\sum_{\terminalqubit i, \lqubit j} \matr{C'}_{\terminalqubit i, \lqubit j} > 0$}
            \State $\mathbf{r}[\terminalqubit i] \gets$
            sum of rows $\terminalqubit i$ in $\matr{C'}$ \Comment{Equation~\ref{eq:row_sum}}
            
            \State $\set A \gets$ $\{\terminalqubit i \}$ rows sums $> 0$, \Comment{ Equation~\ref{eq:available_qubits}} 

            \State $\terminalqubit i \gets $ random element from $\set A$
            
            \State $\set F_{ \textbf{i}}, \matr{C'} \gets$ \Call{\(\function{BestReuseSequence}\)}{$\matr{C'}$, $\terminalqubit i$}
            \Comment{ Algo.~\ref{algo: optimal_reuse_path_for_t}}
            
            \If{$|\set F_{ \textbf{i}}| > 1$}
                \State
                $\set{U'} \gets \set{U'} \| \set F_{ \textbf{i}}$

            \EndIf
        \EndWhile
        
        \State $\set U' \gets$ \Call{$\function{MergeSubsets}$}{$\set U'$}
        \State $\set U' \gets$ \Call{$\function{FinalizeReuse}$}{$\set U'$, $n$}

        \If{$|\set U'| < |\set U|$}
            \State $\set U \gets$ $\set U'$
        \EndIf
        
    \EndFor
    
    \State \Return $\set U$
\end{algorithmic}
\end{algorithm}

Algorithm~\autoref{algo: GidNET} starts by computing the biadjacency and
candidate matrices of the input circuit (line 1)~\cite{fang2023dynamic}. Line 2
initializes the set of reuse sequences \(\set{U}\) with all the qubits in the circuit. Line 3 checks if the candidate matrix is all zeros, indicating an irreducible circuit. If true, line 4 returns \(\set{U}\) with all original qubits, as no reuse is possible.

The core of the algorithm is lines 6-23, where reuse sequences for
\(\graph D\) are computed. Line 6 ensures this is repeated \(\textit{i}\) times,
where \(\textit{i} \propto \log(n)\). The choice of \(\log(n)\) iterations is empirically driven. After evaluating various options, we found using \(\log(n)\) iterations consistently achieved the highest performance across different circuit sizes. Performance was measured based on the likelihood of finding the lowest-width circuit, and the quality of the solution relative to the number of iterations. Using \(\log(n)\) iterations provided an optimal balance, ensuring a high probability of success in finding the best circuit while minimizing computational resources.

Lines 7-8 guarantee that each iteration uses a fresh copy of the circuit's
candidate matrix, \(\matr C'\), and creates a new reuse sequence, \(\set U'
\). Line 9 ensures the algorithm stops when are no more edges in the candidate
matrix. Lines 10 and 11 determine the available qubits from \(\matr C'\) using
~\autoref{eq:available_qubits} and  ~\autoref{eq:row-sum} which are described in Definition~\ref{def:available-qubits}, while line 12 randomly selects a qubit from the available qubits \(\set A\) to compute its reuse sequence. Using Algorithm ~\autoref{algo: optimal_reuse_path_for_t}, line 13 computes the reuse sequence of the selected qubit and updates the candidate matrix to remove one or more edges accordingly.

After identifying the reuse sequence of the selected qubit, line 14 checks if it
is valid (contains more than one qubit). Line 15 appends a valid reuse sequence
to the set of reuse sequences for \(\graph D\). In line 18, the algorithm
consolidates the reuse sequences by merging subsequences with common qubits that could form a single reuse sequence. Line 19 adds any qubits from the original circuit that lack a reuse sequence (ensuring all qubits are accounted for). Lines 20-22 compare the reuse sequence of the previous iteration to the current iteration, ensuring that the best reuse sequences, i.e., the smallest set, is chosen. A smaller set indicates that more qubits are being reused within fewer sequences, effectively reducing the overall width of the original circuit. 

The remainder of this section goes through the details of the most important subroutine in Algorithm~\autoref{algo: GidNET} more carefully.

\subsubsection{\textbf{The $\function{BestReuseSequence}$ Subroutine}}\label{sec: BestReuseSequence}
The $\function{BestReuseSequence}$ subroutine (Algorithm ~\ref{algo:
optimal_reuse_path_for_t}) in line 13 of Algorithm~\autoref{algo: GidNET} computes the reuse sequence, \(\set F_{ \textbf{i}}\), of qubit \(\lqubit {{i}}\) using the candidate matrix, \(\matr C\). 

\begin{algorithm}
\caption{Determine best reuse sequence for qubit \(\lqubit{i}\)}
\label{algo: optimal_reuse_path_for_t}
\begin{algorithmic}[1]
    \Function{BestReuseSequence}{\(\matr{C}\), $\lqubit{i}$}
        \State $\set{F}_{ \textbf{i}} \gets \{\lqubit{i}\}$
        \State $\set{P}_{ \textbf{i}} \gets \{ \lqubit{j} \}$
        \Comment{using Equation~\ref{eq:potential_reuse}}
        \While{$\set{P}_{ \textbf{i}} \neq \emptyset$}
            \State $\dict{D} \gets \{\}$
            \Comment{Placeholder for common neighbors}
            \For{$\lqubit{j} \in \set{P}_{ \textbf{i}}$}
                \State $\dict{D}[\lqubit{j}] \gets \neigbors{j}$
                \Comment{using Equations~\ref{eq:common_neighbors} and \ref{eq:set_T}}
            \EndFor
            \If{$\forall \lqubit{j} \in \dict{D}, \dict{D}[\lqubit{j}] = \emptyset$}
                \State $\lqubit{j} \gets$ random element from $\set{P}_{ \textbf{i}}$
                \State $\set{F}_{ \textbf{i}} \gets \set{F}_{ \textbf{i}} \cup \{\lqubit{j}\}$
                \State $\set{P}_{ \textbf{i}} \gets \dict{D}[\lqubit{j}]$
            \Else
                \State $\dict S[\lqubit{j}] \gets  I_{\textbf{j}}$
                \Comment{using Equations~\ref{eq:max_set}, \ref{eq:max}, and \ref{eq:reuse_score}}

                \State $\lqubit{j} \gets \sim \left\{ \lqubit{j} \mid \dict S[\lqubit{j}] = \max(\dict S), \; \forall \lqubit{j} \in \dict S \right\}$
                \State $\set{F}_{ \textbf{i}} \gets \set{F}_{ \textbf{i}} \cup \{\lqubit{j}\}$
                \State $\set{P}_{ \textbf{i}} \gets \dict{D}[\lqubit{j}]$
            \EndIf
        \EndWhile
        \For{$k = 1$ to $|\set{F}_{ \textbf{i}}| - 1$}
            \State $\terminalqubit{k} \gets \set{F}_{ i}[k]$
            \Comment{terminal node}
            \State $\rootqubit{k+1} \gets \set{F}_{ i}[k+1]$
            \Comment{root node}
            \State $\matr{C} \gets \function{UpdateCMatrix}(\matr{C}, \terminalqubit{k}, \rootqubit{k+1})$
            \Comment{ Algo.~\ref{algo: update_candidate_matrix_after_selection}}
        \EndFor
        \State \Return $\set{F}_{ \textbf{i}}$, \(\matr{C}\)
    \EndFunction
\end{algorithmic}
\end{algorithm}

Examining the relationship between a qubit reuse sequence and the candidate matrix provides insight into $\function{BestReuseSequence}$.
This relationship can be explained by highlighting the edges common to both the candidate graph of the static quantum circuit \(\graph Q\) and the simplified DAG, \(\graph H'\) of the transformed dynamic circuit \(\graph D\). 

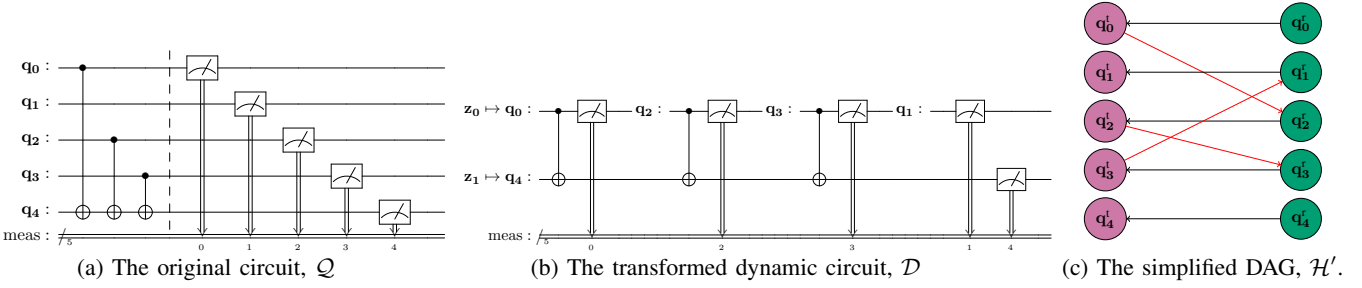
\begin{figure*}[htbp]
    \centering
     \hspace*{2.5mm} 
    \begin{subfigure}{0.25\textwidth}
        \centering
        \scalebox{0.65}{
        \Qcircuit @C=1.0em @R=0.7em {
        & \lstick{\lqubit 0 :  } & \ctrl{4} & \qw & \qw \barrier[0em]{4} & \qw & \meter & \qw & \qw & \qw & \qw & \qw & \qw \\
        & \lstick{\lqubit 1 :  } & \qw & \qw & \qw & \qw & \qw & \meter & \qw & \qw & \qw & \qw & \qw\\
        & \lstick{\lqubit 2 :  } & \qw & \ctrl{2} & \qw & \qw & \qw & \qw & \meter & \qw & \qw & \qw & \qw\\
        & \lstick{\lqubit 3 :  } & \qw & \qw & \ctrl{1} & \qw & \qw & \qw & \qw & \meter & \qw & \qw & \qw\\
        & \lstick{\lqubit 4 :  } & \targ & \targ & \targ & \qw & \qw & \qw & \qw & \qw & \meter & \qw & \qw\\
        & \lstick{\mathrm{meas} :  } & \lstick{/_{_{5}}} \cw & \cw & \cw & \cw & \dstick{_{_{\hspace{0.0em}0}}} \cw \ar @{<=} [-5,0] & \dstick{_{_{\hspace{0.0em}1}}} \cw \ar @{<=} [-4,0] & \dstick{_{_{\hspace{0.0em}2}}} \cw \ar @{<=} [-3,0] & \dstick{_{_{\hspace{0.0em}3}}} \cw \ar @{<=} [-2,0] & \dstick{_{_{\hspace{0.0em}4}}} \cw \ar @{<=} [-1,0] & \cw & \cw\\
        }
        }
        \caption{The original circuit, \(\graph Q\)}
        \label{fig:static_circuit}
    \end{subfigure}%
    \hspace{1.8cm} 
    \begin{subfigure}{0.3\textwidth}
        \centering
        \scalebox{0.6}{
        \Qcircuit @C=0.8em @R=2.9em {
        & \lstick{\vqubit {0} \mapsto \lqubit {0} :} & \ctrl{1} & \meter & \qw & \qw & &   &\lstick{\lqubit {2} :} & \ctrl{1} & \meter & \qw & \qw & & &\lstick{\lqubit {3} :} & \ctrl{1} & \meter & \qw & \qw & & &\lstick{\lqubit {1} :} & \qw & \meter & \qw & \qw & \qw
        \\
        & \lstick{\vqubit {1} \mapsto \lqubit {4} :} & \targ & \qw & \qw & \qw & \qw & \qw & \qw & \targ & \qw & \qw & \qw & \qw & \qw & \qw & \targ & \qw & \qw & \qw & \qw & \qw & \qw & \qw & \qw & \meter & \qw & \qw  
        \\
        & \lstick{\mathrm{meas} :  } & \lstick{/_{_{5}}} \cw & \dstick{_{_{\hspace{0.0em}0}}} \cw \ar @{<=} [-2,0] & \cw & \cw & \cw & \cw & \cw & \cw & \dstick{_{_{\hspace{0.0em}2}}} \cw \ar @{<=} [-2,0] & \cw & \cw & \cw & \cw & \cw & \cw & \dstick{_{_{\hspace{0.0em}3}}} \cw \ar @{<=} [-2,0] & \cw & \cw & \cw & \cw & \cw & \cw & \dstick{_{_{\hspace{0.0em}1}}} \cw \ar @{<=} [-2,0] & \dstick{_{_{\hspace{0.0em}4}}} \cw \ar @{<=} [-1,0] & \cw & \cw\\
        }
        }
        \caption{The transformed dynamic circuit, \(\graph D\)}
        \label{fig:dynamic_circuit}
    \end{subfigure}
    \hspace{1.2cm} 
    \begin{subfigure}{0.24\textwidth}
        \centering
        \scalebox{0.65}{
        \begin{tikzpicture}
            [every join/.style={->, thick, shorten >=1pt},
            node distance=1.5cm, 
            main/.style = {draw, circle}]  
            \foreach \x in {0,...,4}
            {
                \node[main, fill=paleviolet] (s\x) at (0,-\x) {\(\terminalqubit \x\)};
                \node[main, fill=bluegreen] (t\x) at (4,-\x) {\(\rootqubit \x\)};
            }
            \draw[->] (t0) -- (s0);
            \draw[->,red] (s0) -- (t2);
            \draw[->] (t1) -- (s1);
            \draw[->] (t2) -- (s2);
            \draw[->] (t3) -- (s3);
            \draw[->,red] (s2) -- (t3);
            \draw[->,red] (s3) -- (t1);
            \draw[->] (t4) -- (s4);
        \end{tikzpicture}
        }
        \caption{The simplified DAG, \(\graph H'\).}
        \label{fig:dynamic_circuit_graph}
    \end{subfigure}%
    \caption{(a) The original circuit \(\graph Q\) uses 5 logical qubits (\(\lqubit{0} \ldots \lqubit{4}\)), whereas the dynamic circuit \(\graph D\) in (b) uses just 2 virtual qubits (\(\vqubit{0}, \vqubit{1}\)), at the expense of additional depth. (c) The simplified DAG \(\graph H'\) depicts only qubit connections, omitting gate operations. It shows two reuse sequences. The first starts at \(\lqubit{0}\) and passes through \(\lqubit{2}, \lqubit{3}, \lqubit{1}\). The red lines are added to facilitate identification of the qubit reuse sequence. The second is a direct path for \(\lqubit{4}\), with no intermediate qubits.}

    \label{fig:static_and_dynamic_circuits}
\end{figure*}

\myexample{ex:reuse-path-and-candidate-matrix-relationship}{Reuse sequence versus candidate matrix}{
        The simplified DAG \(\graph H'\) in  ~\autoref{fig:dynamic_circuit_graph} has two distinct paths:
        \vspace{-0.5em}
    \begin{enumerate}
        \item \(\rootqubit 0 \rightarrow \terminalqubit 0 \rightarrow \rootqubit 2 \rightarrow \terminalqubit 2 \rightarrow \rootqubit 3 \rightarrow \terminalqubit 3 \rightarrow \rootqubit 1 \rightarrow \terminalqubit 1\)
        \item \(\rootqubit 4 \rightarrow \terminalqubit 4\)
    \end{enumerate}
    
    They correspond to the reuse sequences, \(\set F_{ \textbf{0}}\) and \(\set F_{ \textbf{1}}\) of virtual qubits \(\vqubit {{0}}\) and \(\vqubit {{1}}\) respectively, in the dynamic circuit \(\graph D\). In this representation, each qubit is replaced by a connection from its root to its terminal, indicating when the qubit was initialized and measured, respectively.
    
    In the first path, the edges common to both the candidate graph in ~\autoref{fig:candidate_graph} and \(\graph H'\) in  ~\autoref{fig:dynamic_circuit_graph} are the red-colored edges, \((\terminalqubit 0, \rootqubit 2)\), \( (\terminalqubit 2, \rootqubit 3)\) and \(  (\terminalqubit 3, \rootqubit 1)\), connecting the terminal of one qubit to the root of another qubit in \(\graph H'\). Each edge is an entry in the candidate matrix \(\matr{C}\). The unique set of logical qubits \(\{\lqubit {{0}}, \lqubit {{2}}, \lqubit {{3}}, \lqubit {{1}}\}\) constituting these edges correspond to the qubits in \(\set F_{ \textbf{0}}\), the reuse sequence of virtual qubit \(\vqubit {{0}}\). The second path in \(\graph H'\) has no edge going from the terminal of a qubit to the root of another qubit. Hence, only qubit \(q_{\textbf{4}}\) is in \(\set F_{ \textbf{1}}\), the reuse sequence of virtual qubit \(\vqubit {{1}}\). 
}

Example ~\autoref{ex:reuse-path-and-candidate-matrix-relationship} shows that the set of edges from terminals to roots of any reuse sequence, \(\set F_{ \textbf{i}}\), is a subset of edges in the candidate graph and hence, the candidate matrix. This means that we can determine any \(\set F_{ \textbf{i}}\) from \(\matr{C}\). However, determining the edges that form \(\set F_{ \textbf{i}}\) from the candidate matrix is challenging. Finding an exact solution involves non-convex optimization that would require \(O(n^2)\) variables and \(O(n^4)\) ~\cite{fang2023dynamic, DCKFF22} constraints, making it impractical for large candidate matrices. $\function{BestReuseSequence}$ begins by identifying the sets of qubits that can form a potential reuse sequence in the candidate matrix (line 3 of Algorithm~\autoref{algo: optimal_reuse_path_for_t}).

\mydefinition{def:potential_reuse}{Potential reuse sequence}{
We define  the \textbf{potential reuse sequence}, \(\set P_{ \textbf{i}}\), as the set of logical qubits that can form a reuse sequence with logical qubit \( \lqubit {{i}} \),
\begin{equation}
    \set P_{ \textbf{i}} =  
  \{ \lqubit {{j}} \mid \matr C_{\terminalqubit i, \lqubit j} = 1, \; \forall j \in \{0, 1, \ldots, n-1\} \}.
    \label{eq:potential_reuse}
\end{equation}}

\myexample{ex: potential_reuse_set}{Potential reuse sequence illustration}{
    To find \(\set P_{ \textbf{0}}\) of the candidate matrix in ~\autoref{fig:candidate_matrix}, we examine row \(\terminalqubit 0\) of \(\matr{C}\) and its columns \(\lqubit j\). The columns indexed by  \(\rootqubit 1\), \(\rootqubit 2\), and \(\rootqubit 3\), where the entry in row \(\terminalqubit 0\) is 1, form this set. Thus, \(\set P_{ \textbf{0}} = \{\lqubit 1, \lqubit 2, \lqubit 3 \}\). This indicates that qubits \(\lqubit 1\), \(\lqubit 2\), and \(\lqubit 3\) could each potentially reuse \(\lqubit 0\) after \(\lqubit 0\) completes its operations and could form a potential reuse sequence with \(\lqubit 0\).
}

The set \(\set P_{ \textbf{i}}\) is important because for any reuse sequence,
\(\set F_{ \textbf{j}}\), that begins with logical qubit \(\lqubit {{i}}\), all
other qubits in that sequence must be contained in \(\set P_{
\textbf{i}}\). This is because there is no path in the DAG \(\graph G\), from \( \lqubit {{i}} \) to any \(\lqubit {{j}} \in \set P_{ \textbf{i}} \) in the original circuit.

The next step is to identify qubits in \(\set P_{ \textbf{i}}\) that form the entries in \(\set F_{ \textbf{j}}\) and in what order. The constraint, however, is that the reuse sequence must be acyclic, i.e., no qubit can appear in \(\set F_{ \textbf{j}}\) twice. To ensure this, any qubit \( \lqubit {{k}} \) added to \(\set F_{ \textbf{j}}\) after qubit \( \lqubit {{j}} \) must not have a pre-existing connection in \(\graph G'\) to \( \lqubit {{j}} \), or to any qubit preceding \( \lqubit {{j}} \) in \(\set F_{ \textbf{j}}\). This condition is satisfied if the entry (\(\terminalqubit j, \rootqubit k\)) in the candidate matrix is 1.

Any of the qubits in \(\set P_{ \textbf{i}}\) can be selected to be the first entry of \(\set F_{ \textbf{j}}\). However, to determine which  \(\lqubit j \in \set P_{ \textbf{i}}\) will have the longest reuse sequence while maintaining the acyclicity constraint, we compute the \emph{common neighbors}, \(\neigbors x \) for each  \(\lqubit x \in \set P_{ \textbf{i}}\) and select the qubit whose \(\neigbors x \) has the maximum size, \(|\neigbors x|\).

\mydefinition{def: common_neighbors}{Common Neighbors}{
    We define \(\neigbors x \), the \textbf{common neighbors} of qubit \(\lqubit x\) as: 
\begin{eqnarray}
    \neigbors x = \bigcap_{\lqubit k \in \set T_{\textbf{x}}} \set P_{ \textbf{k}}, 
\label{eq:common_neighbors}
\end{eqnarray}
\noindent where
\begin{eqnarray}
    \set T_{\textbf{x}} = \{ \set F_{ \textbf{j}} \cup \{\lqubit x\} \}
    \label{eq:set_T}
\end{eqnarray}
is the \emph{updated reuse sequence}.
}

The \(\neigbors x \) with the maximum size is chosen because it contains more qubits to select from, increasing the likelihood of finding the longest path.

\myexample{ex: compute_reuse_path}{Common neighbors illustration}{
    Recall that in ~\cref{ex: potential_reuse_set}, we computed \(\set P_{ \textbf{0}} = \{\lqubit 1, \lqubit 2, \lqubit 3 \}\). To determine the optimized \(\set F_{\textbf{0}}\) that could be derived from \(\set P_{ \textbf{0}}\), we
    
    \begin{enumerate}
        \item Initialize \(\set F_{\textbf{0}} = \{\lqubit{0}\}\)
        \item Using ~\autoref{eq:potential_reuse} compute \(\set P_{ \textbf{i}}\) for each \(\lqubit x \in \set P_{ \textbf{0}}\) as: \(\set P_{ \textbf{1}} = \{\lqubit 0, \lqubit 2, \lqubit 3, \lqubit 4 \}\), \(\set P_{ \textbf{2}} = \{\lqubit 1, \lqubit 3 \}\) and \(\set P_{ \textbf{3}} = \{\lqubit 1 \}\)

        \item Using~\autoref{eq:common_neighbors} and ~\autoref{eq:potential_reuse}, compute \(\neigbors x \) for each \(\lqubit x \in \set P_{ \textbf{0}}\):
        \begin{itemize}
            \item For \(\lqubit 1\): 
            \vspace{-1.0em}
            \begin{eqnarray*}
                \set T_{\textbf{1}} &=& \set F_{ \textbf{0}} \cup \{\lqubit 1\} = \{\lqubit 0, \lqubit 1\},\\
                \neigbors 1 &=& \set P_{ \textbf{0}} \cap \set P_{ \textbf{1}}
                = \{\lqubit 2, \lqubit 3 \}.
            \end{eqnarray*}
            \vspace{-2.0em}
            \item For \(\lqubit 2\): 
            \vspace{-1.0em}
            \begin{eqnarray*}
                \set T_{\textbf{2}} &=& \set F_{ \textbf{0}} \cup \{\lqubit 2\} = \{\lqubit 0, \lqubit 2\},\\
                \neigbors 2 &=& \set P_{ \textbf{0}} \cap \set P_{ \textbf{2}} = \{\lqubit 1, \lqubit 3 \}.
            \end{eqnarray*}
            \vspace{-2.0em}
            \item For \(\lqubit 3\):
            \vspace{-1.0em}
            \begin{eqnarray*}
                \set T_{\textbf{3}} &=& \set F_{ \textbf{0}} \cup \{\lqubit 3\} = \{\lqubit 0, \lqubit 3\},\\
                \neigbors 3 &=& \set P_{ \textbf{0}} \cap \set P_{ \textbf{3}} 
                = \{\lqubit 1 \}.
            \end{eqnarray*}
        \end{itemize}
        \item Compute the size of each \(\neigbors x \):
        \(|\neigbors 1 | = |\neigbors 2 | = 2 \), \(|\neigbors 3 | = 1 \).
    \end{enumerate} 
}

Example~\autoref{ex: compute_reuse_path} shows multiples \(\neigbors x\) can
have the same maximum size. One option is to randomly select one of them, but
this may result in sub-optimal reuse sequences. Instead, we define a secondary quality metric, which we denote as the \emph{reuse score}.

\mydefinition{eq:reuse_score}{Reuse score}{
    We define the \emph{reuse score}, \(I_{\textbf{j}}\), of qubit \(\lqubit j\) as:
\vspace{-0.7em}
\begin{eqnarray}
    I_{\textbf{j}} =  \sum_{\left\{ \textbf{j} \neq \textbf{k}, \; \forall  \lqubit{j} \in \set M \right \}} { \left| \neigbors j \cap \neigbors k \right|},
\end{eqnarray}
\vspace{-1.0em}
where
\vspace{-0.8em}
\begin{eqnarray}
    \set M = \left\{\lqubit j \mid     \left| \neigbors j\right| = s, \; \forall \lqubit j \in \set P_{ \textbf{i}} \right\},    \label{eq:max_set}
\end{eqnarray}
\vspace{-1.0em}
and
\vspace{-1.0em}
\begin{eqnarray}
    s = \max \left(\left \{\left | \neigbors j \right |, \; \forall \lqubit j \in \set P_{ \textbf{i}} \right \} \right) 
    \label{eq:max}
\end{eqnarray}
is the maximum size of any set of common neighbors.}

The qubit with the highest \(I_{\textbf{j}}\) is chosen because it has a greater
influence on the reuse sequence, \(\set F_{\textbf{j}}\), compared to other
qubits. Here, influence means \(\lqubit j\) has greater potential to further extend
\(\set F_{\textbf{j}}\) if chosen, because of its \(\neigbors x
\). If multiple qubits have the same \(I_{\textbf{j}}\), one of them is selected
at random. 

\myexample{ex: reuse_score}{Reuse score illustration}{
    Given that \(|\neigbors 1| = |\neigbors 2|\), we compute  \(I_{\textbf{j}}\) for each qubit,\\ 
\begin{equation*}
    I_{\textbf{1}} = \sum \left\{ \left| \neigbors 1 \cap \neigbors 2 \right |
\right\}  = 1, \quad
    I_{\textbf{2}} = \sum \left\{ \left| \neigbors 2 \cap \neigbors 1 \right | \right\} = 1.
\end{equation*}

\noindent Since both qubits have the same reuse score, we can
randomly choose one to extend \(\set F_{ \textbf{0}}\). Suppose \(\lqubit 2\) is selected. Then 
 \[\set F_{\textbf{0}} = \set T_{\textbf{2}} = \{\lqubit 0, \lqubit 2\}.\]
}

For every qubit \(\lqubit j\) selected from \(\set P_{ \textbf{i}}\) and appended to \(\set F_{\textbf{i}}\), the \(\neigbors x \) of the selected qubit \(\lqubit j\) becomes the updated  \(\set P_{ \textbf{i}}\). The set \(\set P_{ \textbf{i}}\) is iteratively updated with \(\neigbors x \) after each selection of \(\lqubit j\) from \(\set P_{ \textbf{i}}\) until no further qubits can be added to \(\set F_{ \textbf{i}}\) from \(\set P_{ \textbf{i}}\).

\myexample{ex:compute_reuse_path_2}{Reuse sequence computation}{
    Having determined \(\set F_{\textbf{0}} = \{\lqubit 0, \lqubit 2\}\), and
\(\set P_{ \textbf{0}}\) is updated to \(\set P_{ \textbf{0}} = \neigbors 2 = \left\{\lqubit 1, \lqubit 3\right\} \), we want to select the next qubit from the updated \(\set P_{ \textbf{0}}\). Following the steps in ~\cref{ex: compute_reuse_path}, we have:

    \begin{enumerate}
        \item Using equations~\autoref{eq:common_neighbors} and
~\autoref{eq:potential_reuse}, we compute \(\neigbors x \) for each \(\lqubit x \in \set P_{ \textbf{0}}\)
        \vspace{-0.3em}
        \begin{itemize}
            \item For \(\lqubit 1\):
            \vspace{-0.9em}
            \begin{eqnarray*}
                \set T_{\textbf{1}} &=& \set F_{ \textbf{0}} \cup \{\lqubit 1\} = \{\lqubit 0, \lqubit 2, \lqubit 1\},\\
                \neigbors 1 &=&  \set P_{ \textbf{0}} \cap \set P_{ \textbf{2}} \cap \set P_{ \textbf{1}} 
                = \{\lqubit 3 \}.
            \end{eqnarray*}
            \vspace{-2.0em}
            \item For \(\lqubit 3\): 
            \vspace{-0.9em}
            \begin{eqnarray*}
                \set T_{\textbf{3}} &=& \set F_{ \textbf{0}} \cup \{\lqubit 3\} = \{\lqubit 0, \lqubit 2, \lqubit 3\},\\
                \neigbors 3 &=& \set P_{ \textbf{0}} \cap \set P_{ \textbf{2}} \cap \set P_{ \textbf{3}} = \{\lqubit 1 \}.
            \end{eqnarray*}
            \vspace{-0.9em}
        \end{itemize}
        \vspace{-0.9em}
        \item compute the size of each \(\neigbors x \):
        \(|\neigbors 1 | = 1 \) and \(|\neigbors 3 | = 1 \).
    \end{enumerate} 
Both \(\lqubit 1\) and \(\lqubit 3\) have the same \(|\neigbors x | =
1\). Computing \(I_{\textbf{j}}\) for each yields \(I_{\textbf{1}} = I_{\textbf{3}} = 0\). This means we can randomly select either  \(\lqubit 1\) or \(\lqubit 3\) to append to \(\set F_{ \textbf{0}}\). 
Suppose \(\lqubit 3\) is randomly selected, resulting in \(\set F_{ \textbf{0}}
= \{\lqubit 0, \lqubit 2, \lqubit 3\}\), then the updated \(\set P_{ \textbf{0}}
= \neigbors 3 = \{\lqubit 1\}\). Since there is only one element left in \(\set P_{ \textbf{0}}\), we append it to \(\set F_{ \textbf{0}}\) and obtain \(\set F_{ \textbf{0}} = \{\lqubit 0, \lqubit 2, \lqubit 3, \lqubit 1\}\) and \(\set P_{ \textbf{0}} = \emptyset\). This completes the reuse sequence \(\set F_{ \textbf{0}}\) of qubit \(\lqubit {{0}}\).
}

\begin{algorithm}
\caption{Update Candidate Matrix After Selection}
\label{algo: update_candidate_matrix_after_selection}
\begin{algorithmic}[1]
\Procedure{UpdateCMatrix}{\(\matr{C}\), $\terminalqubit i$ , $\lqubit j$}
    \State $\set V_r \gets \{\rootqubit k \,|\, \matr{C}_{\terminalqubit i, \rootqubit k} = 0, \; \forall k \in \{0, 1, \ldots, n-1\} \}$ 
    \State $\set V_t \gets \{\terminalqubit k \,|\, \matr{C}_{\terminalqubit k , \lqubit j} = 0, \; \forall k \in \{0, 1, \ldots, n-1\}\}$ 
    
    \For{$(\terminalqubit k, \rootqubit l) \in \{\set{Q}_t\} \times \{\set{Q}_r\}$}
        \State $\matr{C}_{\terminalqubit k, \rootqubit l} \gets 0$ 
    \EndFor
    \State $\matr{C}_{\terminalqubit i, \rootqubit k} \gets 0, \; \forall k \in \{0, 1, \ldots, n-1\}$ 
    \State $\matr{C}_{\terminalqubit k, \lqubit j} \gets 0, \; \forall k \in \{0, 1, \ldots, n-1\}$
    \State \Return \(\matr{C}\) 
\EndProcedure
\end{algorithmic}
\end{algorithm}

After the reuse sequence of a qubit is determined, the
\(\function{UpdateCMatrix}\) procedure in algorithm ~\autoref{algo:
update_candidate_matrix_after_selection} updates the candidate matrix, as done
in ~\cite{fang2023dynamic}. This is crucial to avoid multiple selections of an edge. It also prevents the creation of cycles in the DAG and eliminates potential conflicts in qubit reuse.
It is possible that edges remain in the candidate matrix after
the update; this signals more qubit reuse opportunities. In this case, a new set of available qubits is derived from the updated candidate matrix.

\mydefinition{def:available-qubits}{Available qubits}{
    We define the available qubits in the candidate matrix as follows:
    \begin{eqnarray}
    \label{eq:available-qubits}
        \set{A} &=& \{\lqubit i \mid \mathbf{r}[\lqubit i] > 0, \; \forall \text{i} \in \{0, 1, \ldots, n-1\} \} \label{eq:available_qubits}
    \end{eqnarray}
    where $\mathbf{r}$ is the sum of each row in $\matr C$,
    \begin{eqnarray}
    \label{eq:row-sum}
        \mathbf{r}[\lqubit i] &=& \sum_{\textbf{j}=0}^{n-1} \matr{C'}_{\terminalqubit i, \lqubit j} \quad \forall \text{i} \in \{0, 1, \ldots, n-1\} \label{eq:row_sum}. 
    \end{eqnarray}
}
\vspace{-0.7em}
Upon confirming availability of additional qubits, another qubit is selected,
its reuse sequence computed, and the candidate matrix updated. This process is
repeated until no edges remain in the candidate matrix (line 23 of Algorithm ~\autoref{algo: optimal_reuse_path_for_t}).

\subsection{Complexity Analysis of the GidNET Qubit Reuse Algorithm}
\label{subsec:complexity_analysis_oz_gidnet}
The time complexity of initializing variables and computing matrices for a
quantum circuit depends on its size. Given a circuit with \(n\) qubits and \(m\)
gates, it is converted into a DAG in \(O(m)\) time, then simplified using Depth
First Search (DFS). Running DFS to determine an edge \((\rootqubit j, \terminalqubit i)\) in the biadjacency graph \(\graph G'\) has complexity \(O(m)\). In \(\graph G'\), there are at most \(n^2\) edges from roots \(\rootqubit j\) to terminals \(\terminalqubit i\), resulting in an overall complexity of \(O(mn^2)\). Computing the biadjacency matrix from \(\graph G'\), and then the candidate matrix, both take \(O(n^2)\). 

After determining the optimized reuse sequence \(\set U\), the DAG is converted to a dynamic circuit. For dynamic circuit conversion, edges are added to the DAG from \(\set U\), requiring \(O(m)\) for topological sorting and another \(O(m)\) for appending instructions. The final step of updating instructions for each qubit takes \(O(mn)\), resulting in a total time complexity of \(O(m) + O(m) + O(mn) = O(mn)\). Hence, the total time complexity for data structure creation and postprocessing is \(O(m) + O(n^2) + O(mn^2) +  O(mn) = O(mn^2)\).

Next, we examine lines 6-17 of Algorithm ~\autoref{algo: GidNET}. For each of
the \(\log n\) iterations (line 6), the algorithm iterates over the available
qubits \(\set{A}\) in the candidate matrix (line 11) to randomly select a qubit
(line 12), leading to an outer loop complexity that scales as \(O(\log n)\). At
each iteration, the candidate matrix is duplicated to preserve its original
state for subsequent evaluations (line 7); this involves \(O(n^2)\) operations
to replicate each matrix element. A significant portion of the algorithm is
spent in a while loop (lines 9-17), determining potential reuse sequences (see
~\autoref{sec: optimal_reuse_path_for_t}) until exhaustion. The sum operation
within this loop, necessary to evaluate the continuation condition (line 9), exhibits \(O(n^2)\) complexity per iteration. The progressive reduction of viable qubits 
reduces the computational load of subsequent iterations.

The invocation of \(\function{BestReuseSequence}{}\) ~\autoref{algo:
optimal_reuse_path_for_t} (line 13) constitutes the algorithmic core, tasked
with identifying the most efficient reuse sequence for a selected qubit. This
function's complexity, preliminarily estimated at \(O(n^3)\) (see
~\autoref{sec:complexity_oz_optimal_t_reuse_path}), dominates the overall
computational effort, attributed to extensive matrix operations and sequence updates. Assuming the main loop iterates \(p\) times, where \(p < n\), the cumulative complexity approximates to \(O(p \cdot n^3)\). Post-calculation, the algorithm merges overlapping sequences and adjusts the reuse list to include all qubits.  \(\function{MergeSubsets}{}\)  (line 18) merges pairs of elements in the final reuse sequence into interconnected subsequences, ensuring all elements connected directly or indirectly are in the same subsequence. This involves iteratively checking and merging pairs with a function that finds a common subsequence for a given pair, and another function that merges two subsequences while maintaining their order. The merging process, including additional passes to ensure all interconnected subsequences are fully merged, yields a time complexity of \(O(\mu \times \nu \times k^2)\), where \(\mu\) is the number of pairs, \(\nu\) is the number of merged subsequences, and \(k\) is the average length of each subsequence. 

 \(\Call{FinalizeReuse}{}\)  (line 19) ensures all qubits are accounted for. It
first creates a set of all qubits in the current reuse sequence, then compares
this to a set of the original qubits (\(n\)) to find any missing qubits.
Missing qubits are appended as individual elements to the set of reuse
sequences. This guarantees that every qubit is present, either as part of a
reuse sequence or independently, resulting in a complete and finalized qubit reuse list. The time complexity is \(O(n)\).

Considering the algorithm's iterative nature and the significant impact of converting the circuit DAG to simplified DAG, the \(\function{BestReuseSequence}{}\) function, and \(\function{MergeSubsets}{}\) the cumulative complexity of GidNET can be conservatively approximated as \(O(mn^2 + \mu \nu k^2 + pn^3\log{n})\).

\subsubsection{Complexity Analysis of BestReuseSequence Function}
\label{sec:complexity_oz_optimal_t_reuse_path}

The \(\function{BestReuseSequence}\) function (Algorithm ~\autoref{algo:
optimal_reuse_path_for_t}) is pivotal in determining the optimal reuse sequence
for a qubit \(\terminalqubit i\). Initialization of \(\set F_{ \textbf{i}}\)
(line 2) and \(\set P_{ \textbf{i}}\) (line 3) are \(O(1)\) and \(O(n)\)
operations respectively. In lines 4-19, the while loop iterates over potential
qubits in \(\set P_{ \textbf{i}}\) until no further qubits can be added to
\(\set F_{ \textbf{i}}\), potentially engaging with all \(n\) qubits in \(\set
P_{ \textbf{i}}\). Inside the while loop, for each qubit, \(\lqubit j\) in the
current \(\set P_{ \textbf{i}}\), \(\neigbors j\) is computed and stored in the
placeholder \(\dict D\). Each \(\neigbors j\) requires \(O(n)\) complexity. Thus, the for loop runs in \(O(n) \times O(n) = O(n^2)\) to compute all \(\neigbors j\). 

Lines 9-12 handle the case where all computed \(\neigbors j\) in \(\dict D\) are
empty, which involves checking if all values in \(\dict D\) are empty, in time \(O(n)\). Lines 13-18 are the cases where at least one \( \neigbors j\) is not empty and multiple qubits, \(\lqubit j\) have maximum \( \neigbors j\). Finding the placeholder \(\dict S\) in line 14 using Equations~\ref{eq:max_set}, \ref{eq:max}, and \ref{eq:reuse_score} involves iterating over each \(\lqubit j\) in \(\set M\). During each iteration, it computes intersections with \(O(n)\) complexity, calculates the sum of intersections with \(O(n)\) complexity and updates \(\dict S\) with \(O(1)\) complexity. This results in a total complexity of \(O(n) \times (O(n) + O(n) + O(1)) = O(n^2)\). Line 15 has \(O(n)\) complexity, while each update in lines 16-17 requires \(O(1)\) complexity. Hence, the dominant time complexity of lines 9-18 is \(O(n^2)\).

Finally, lines 20-24 call
\(\function{UpdateCMatrix}\) to update \(\matr C\) if \(|\set F_{ \textbf{i}}| > 1\). Each call to \(\function{UpdateCMatrix}\) is \(O(n^2)\) (see Algorithm ~\autoref{algo: update_candidate_matrix_after_selection} in ~\autoref{sec: complexity_oz_update_candidate_matrix_after_selection}), and the loop runs up to \(n\) times giving a total complexity of \(O(n^2) \times O(n) = O(n^3)\). Thus, the overall time complexity of the \(\function{BestReuseSequence}\) is \(O(n^2) + O(n^3) + O(n^2) + O(n^3) = O(n^3)\).

\subsubsection{Complexity Analysis of Update Candidate Matrix After Selection}\label{sec: complexity_oz_update_candidate_matrix_after_selection}

The function updates the matrix based on a chosen qubit's terminal
\(\terminalqubit i\) and another qubit's root \(\rootqubit
j\). 
Identifying nodes not directly connected to \(\terminalqubit i\) and \(\lqubit
j\) requires examining each element along a specified row and column, which is
\(O(n)\) overall. Setting an entire row and column to zero is also \(O(n)\). The
most intensive task is the nested loop that runs for all pairs in the Cartesian
product of the set of root vertices \(\set V_r\), and the set of terminal vertices \(\set V_t\) (lines 4-6), updating the matrix to account for indirect connections and possibly requiring a review and update of every matrix entry. This step dominates, leading to an overall complexity of \(O(n^2)\).

\section{Experimental Setup and Results} \label{sec:experimental_results}
In this section, we present the experimental setup, benchmarking results, and the statistical methods used to validate the performance improvements of GidNET.

Two qubit reuse algorithms were compared against GidNET: QNET ~\cite{fang2023dynamic}, and the Qiskit implementation ~\cite{QiskitQubitReuse2023} of the qubit reuse algorithm in ~\cite{DCKFF22}. Due to the proprietary implementation of  ~\cite{DCKFF22}, direct access was unavailable as it requires a subscription to Quantinuum’s H-series hardware \footnote{M. DeCross, E. Chertkov, M. Kohagen, and M. Foss-Feig, personal communication, April 18, 2024.}. The authors provided detailed insights into the algorithm's principles for academic understanding. For our study, we assume the Qiskit implementation reflects the original algorithm's functionality.

Experimental validation of GidNET was conducted on a set of benchmark circuits, including both Google random circuit sampling (GRCS) and Quantum Approximate Optimization Algorithm (QAOA) circuits. We specifically chose these circuits because they were used in prior works to assess qubit reuse algorithms on near-term quantum devices~\cite{DCKFF22, fang2023dynamic}.

Benchmarks were conducted on a virtual machine hosted by an Acer Nitro N50-640 desktop running Windows 11 x64. The host system was equipped with a 12th Gen Intel(R) Core(TM) i7-12700F 2.10 GHz processor and 16 GB of RAM. Virtualization was facilitated by VMware Workstation 17, which ran an Ubuntu 22.04.1 LTS virtual machine. The VM was configured with 8 CPU cores, 16 GB of RAM. 

To assess the runtime efficiency of GidNET, QNET, and Qiskit, \texttt{"\%timeit -o"} command was employed within a JupyterLab notebook environment. While it is acknowledged that high-performance computing resources could potentially enhance the feasibility range of the tested algorithms, the results presented herein aim to approximate the conditions accessible to most researchers and developers.

To ensure consistency, each randomly generated circuit was seeded such that each algorithm operated on identical circuit configurations. 
Performance data collected from the application of the three algorithms was analyzed to determine their efficacy. Runtime and circuit width before and after application of each algorithm were recorded. Statistical tests using polynomial regression were employed to validate the theoretically obtained complexity analyses of GidNET and QNET against experimental
results.

\subsection{Google Random Circuit Sampling (GRCS) }
GRCS was introduced to demonstrate quantum computers' capability to solve problems intractable for classical computers, using random quantum circuits ~\cite{boixo2018characterizing}. GRCS utilizes circuits designed for qubits arranged in an \(n_1 \times n_2\) lattice, which incorporates several cycles of quantum gates on all qubits. First, a layer of Hadamard gates is applied across all qubits. Subsequent cycles include a structured pattern: a layer of controlled \(Z\) (\(CZ\)) alternately arranged according to some pre-defined patterns, succeeded by a layer of single-qubit gates—\(\{X^{1/2}, Y^{1/2}, T\}\)—targeted at qubits not involved in the \(CZ\) gates during the same cycle.

GRCS circuits with 16 to 144 qubits were generated at depths 11, 12, and 15 for each qubit number. The performance of the algorithms was evaluated based the reduction in circuit width and the overall runtime.

Due to the random nature of GidNET and QNET, each GRCS circuit was benchmarked 10 times, and the best (or smallest) compiled circuit width recorded. Qiskit was only run once for each circuit since it is a deterministic algorithm. 

\begin{figure}[ht]
  \centering
  \includegraphics[width=\linewidth]{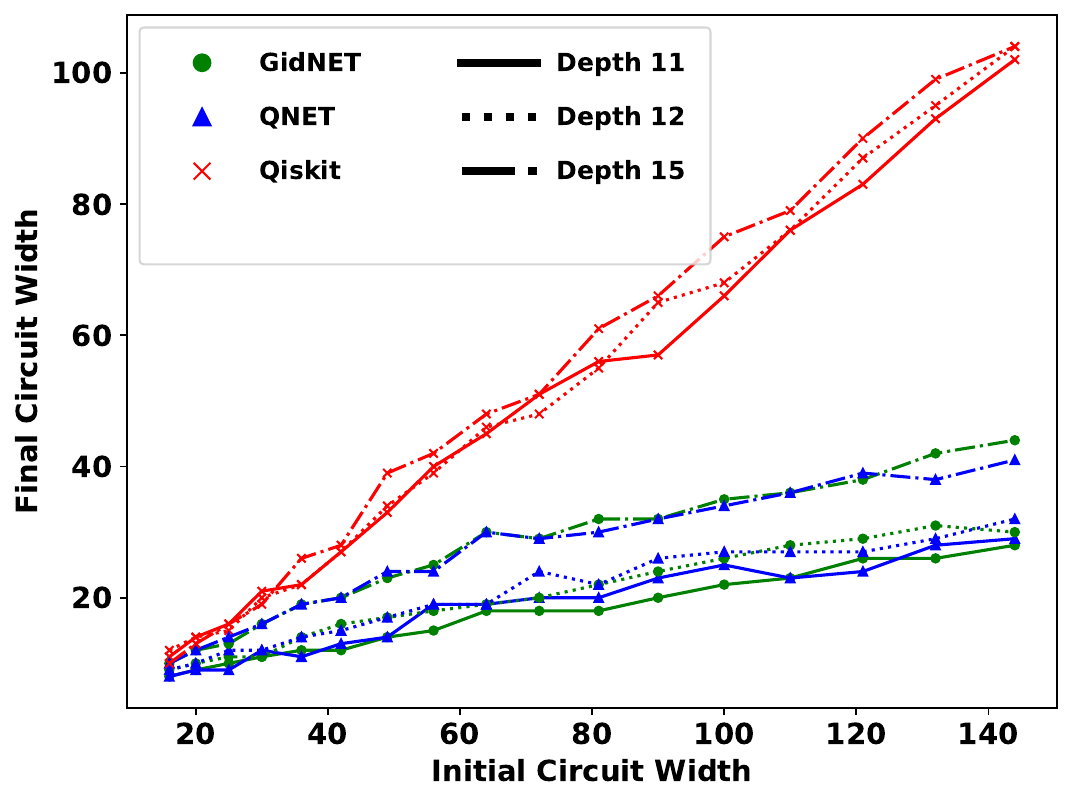}
  \caption{Comparison of the final circuit width achieved after qubit reuse compilation by GidNET, QNET, and Qiskit algorithms for GRCS circuits. GidNET performed better than the other algorithms in most cases.}
  \label{fig:supremacy_circuit_width_comparison}
\end{figure}

~\autoref{fig:supremacy_circuit_width_comparison} shows the final circuit
widths. GidNET achieved an improved circuit width for depth 11 GRCS circuits by
an average of 4.4\% (the geometric mean is used for averages reported throughout), with reductions reaching up to 21\% for circuit size 56 compared to QNET. Also, GidNET offers a substantial advantage in width reduction over Qiskit, outperforming it by an average of 59.3\%. This improvement becomes increasingly significant for larger circuits, reaching up to 72\% width reduction for circuit sizes 132 and 144. For depths 12 and 15, GidNET and QNET achieved similar performance, but they consistently outperformed Qiskit.

\begin{figure}[ht]
  \centering
  \includegraphics[width=\linewidth]{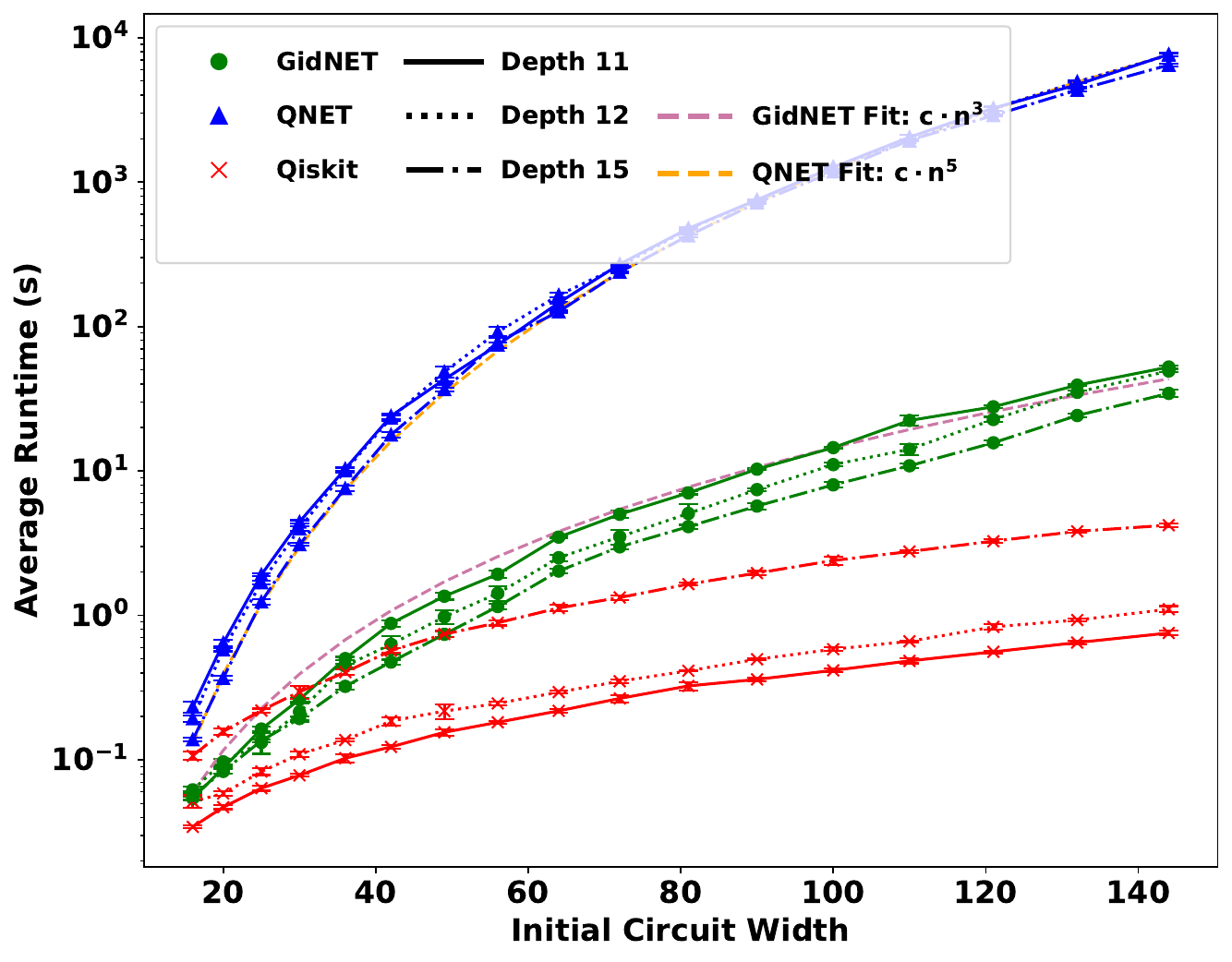}
  \caption{Comparison of GidNET, QNET, and Qiskit average runtimes for GRCS circuits.}
  \label{fig:supremacy_runtime_comparison}
\end{figure}

To determine runtime performance, each circuit configuration was run a total of 7 times. The average runtime from these executions was then computed, and is shown in ~\autoref{fig:supremacy_runtime_comparison}.

Our study demonstrates that for GRCS circuits of depth 11, GidNET is faster by
an average of 97.4\% (or 38.5$\times$) reaching up to 99.3\% (or 142.9$\times$) for circuit size 144 compared to QNET. This result shows GidNET's superior ability to optimize quantum circuits, evidencing not only reduced circuit width but also enhanced runtime performance. Although Qiskit is on average 92.1\% (or 12.7$\times$) faster in runtime, the significant improvement in circuit width (59.3\%) by GidNET can be considered an acceptable tradeoff. For GRCS circuits of depths 12 and 15, GidNET is on average 46.9$\times$ and 48.7$\times$ faster than QNET respectively. Although Qiskit is 7.3$\times$ and 1.7$\times$ faster for GRCS circuits of depths 12 and 15 respectively, GidNET achieved a much better average width reduction (53.6\% and 40.0\%, respectively).

\subsection{Quantum Approximate Optimization Algorithm (QAOA) }
The second benchmark set comprised the Quantum Approximate Optimization Algorithm (QAOA) ~\cite{farhi2014quantum, DCKFF22, fang2023dynamic}, which is designed for solving combinatorial optimization problems on quantum computers. The algorithm operates by applying a unitary transformation \(U(\vec{\beta}, \vec{\gamma})\), structured as a product of alternating operators, which are iteratively optimized to minimize the cost function encoded in the quantum circuit. The core of the QAOA is formed by repeatedly applying two types of unitary transformations for a total of \( p \) layers, where \( p \) influences solution quality and computational complexity:
\begin{equation}
U(\vec{\beta}, \vec{\gamma}) = \prod_{n=1}^p U_B(\beta_n) U_C(\gamma_n). \label{eq:unitary}
\end{equation}
Here, \( U_B(\beta_n) = e^{-i\beta_n H_B} \) and \( H_B = \sum_i X_i \) is the mixing Hamiltonian, and \( U_C(\gamma_n) = e^{-i\gamma_n H_C} \) encodes the problem's cost Hamiltonian. We used that of MaxCut,
\begin{equation}
H_C = \frac{1}{2} \sum_{(i,j) \in E} (1 - z_i z_j). \label{eq:cost_hamiltonian}
\end{equation}
We explored the runtime efficiency, and solution quality (with respect to the final or compressed circuit width) of applying the three algorithms to QAOA MaxCut problems on random unweighted three-regular (U3R) graphs. An U3R graph is a type of graph in which each vertex is connected to exactly three other vertices~ ~\cite{chai2022shortcuts}. The structural simplicity of U3R graphs, where the number of quantum gates mirrors the number of edges and scales linearly with the number of vertices, makes it a standard benchmark for QAOA on current hardware ~\cite{zhou2020quantum, harrigan2021quantum, ebadi2022quantum}. Their shallow, broad layout, coupled with sparse gate connectivity, position MaxCut QAOA circuits as an ideal testbed for qubit reuse strategies.

\begin{figure}[ht]
  \centering
  \includegraphics[width=\linewidth]{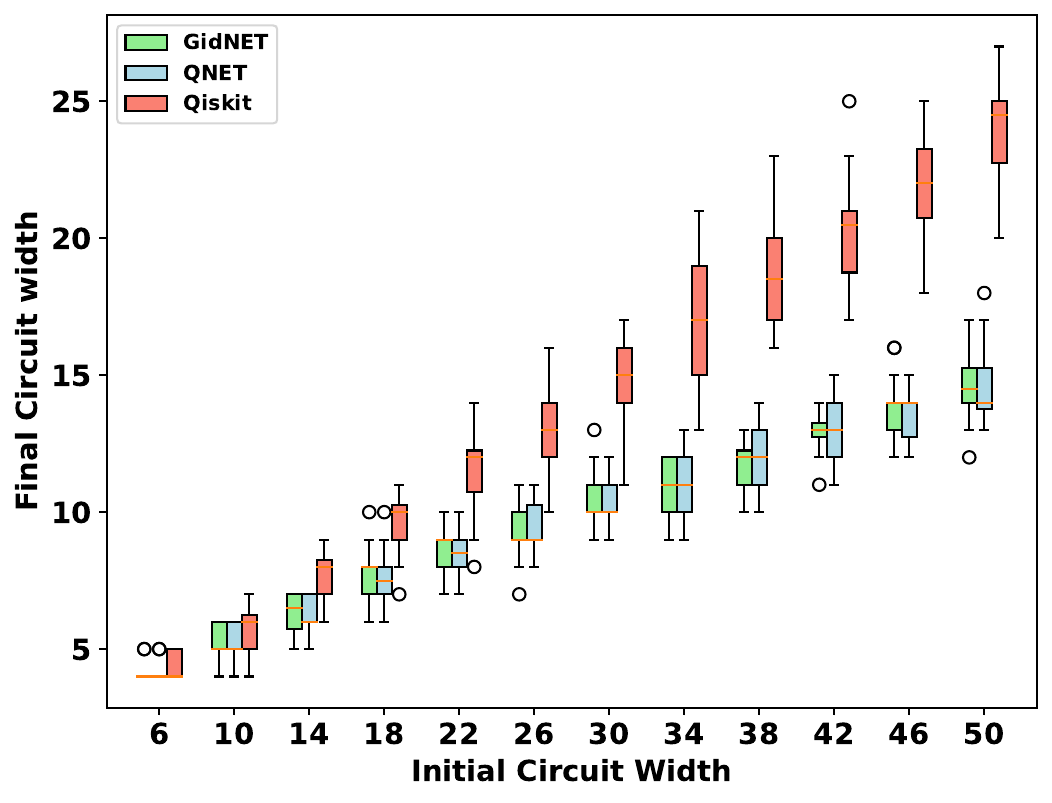}
  \caption{Comparison of final circuit widths achieved after qubit reuse compilation by GidNET, QNET, and Qiskit for QAOA circuits, \(p=1\).}
  \label{fig:qaoa_width_comparison_p1_boxplot}
\end{figure}

\begin{figure}[ht]
  \centering
  \includegraphics[width=\linewidth]{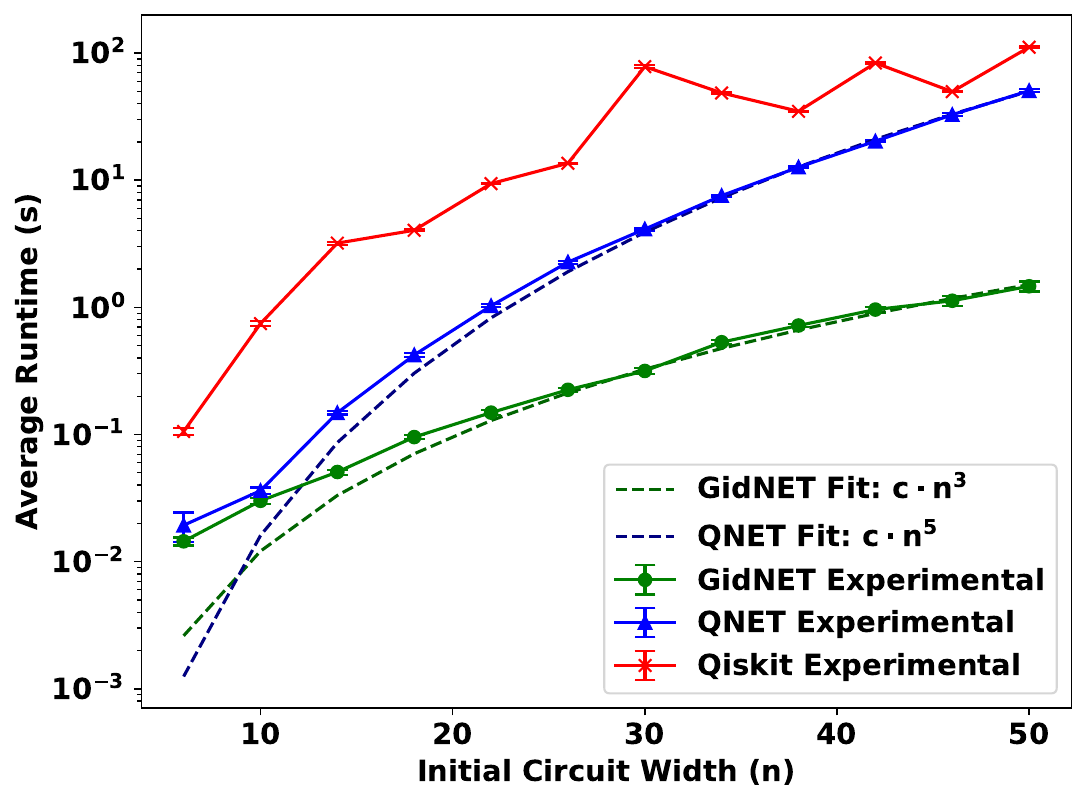}
  \caption{Comparison of GidNET, QNET and Qiskit Average Runtime for QAOA Circuit, \(p\)=1.}
  \label{fig:qaoa_circuit_runtime_p1}
\end{figure}

\begin{figure}[ht]
  \centering
  \includegraphics[width=\linewidth]{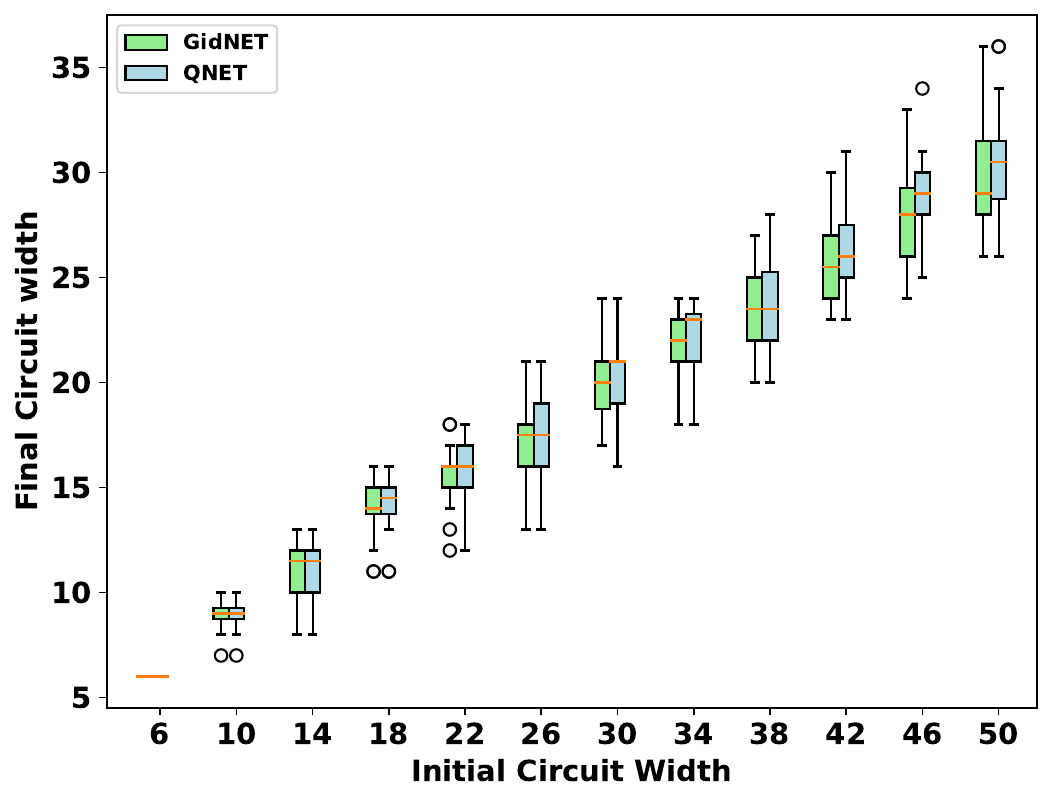}
  \caption{Comparison of final circuit widths after qubit reuse compilation by GidNET and QNET for QAOA circuits, \(p\)=2.}
  \label{fig:qaoa_circuit_width_p2_boxplot}
\end{figure}

\begin{figure}[ht]
  \centering
  \includegraphics[width=\linewidth]{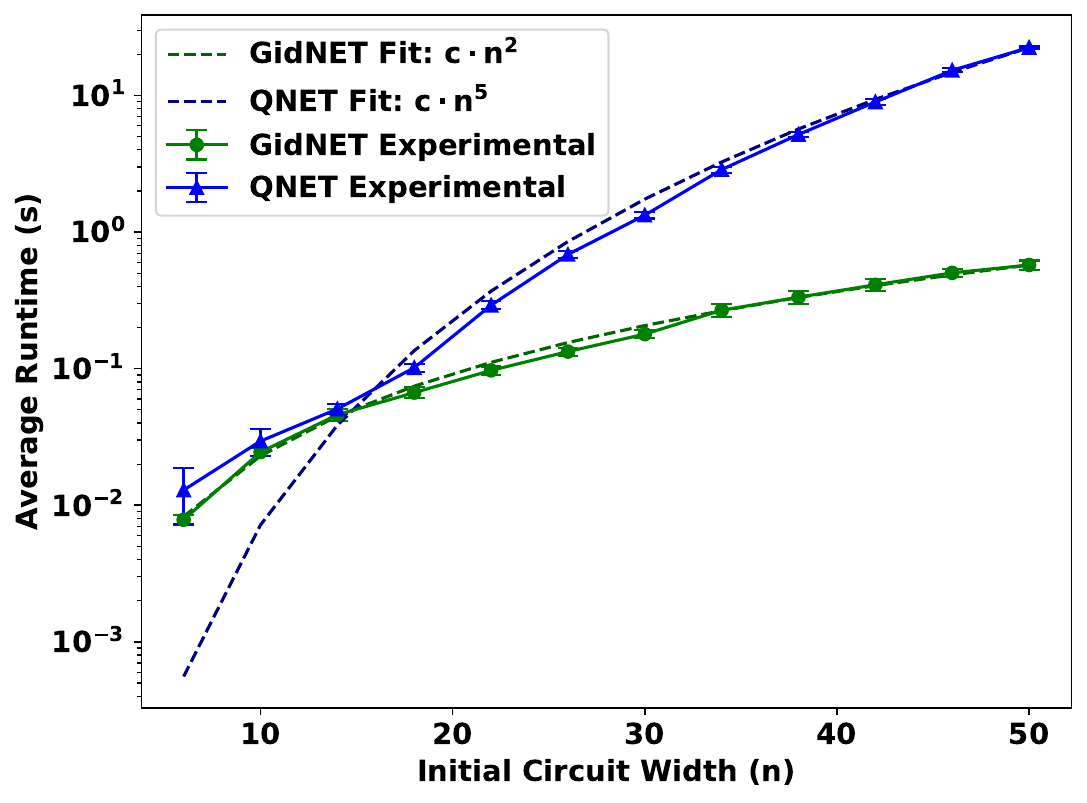}
  \caption{Comparison of GidNET and QNET Average Runtime for QAOA Circuit, \(p\)=2 (average over 7 runs of 1 graph).}
  \label{fig:qaoa_circuit_runtime_p2}
\end{figure}

Each algorithm was tested on QAOA circuits for two depths, \( p = 1 \) (~\autoref{fig:qaoa_width_comparison_p1_boxplot} and ~\autoref{fig:qaoa_circuit_runtime_p1}) and \( p = 2 \) (~\autoref{fig:qaoa_circuit_width_p2_boxplot} and ~\autoref{fig:qaoa_circuit_runtime_p2}), to evaluate how depth affects compression and performance for varying number of qubits. For each fixed number of qubits, 20 random U3R graphs were generated using the NetworkX package ~\cite{hagberg2008exploring}. The same set of random graphs were utilized to construct QAOA circuits for each of the three competing qubit reuse algorithms. Both GidNET and QNET algorithms were run 10 times for each circuit generated from each random graph, and the best result was recorded.

~\autoref{fig:qaoa_width_comparison_p1_boxplot} shows the boxplot comparing the circuit widths achieved by GidNET, QNET, and Qiskit for various circuit sizes.  The range of widths, indicated by the vertical span of the boxes and whiskers, provides insight into the spread and median values of the circuit widths for each qubit reuse framework's performance for different circuit sizes.
Outliers are shown as individual points beyond the whiskers. The presence of outliers can indicate specific instances where a framework handles certain circuits unusually well (or poorly).

GidNET and QNET show similar performance in terms of circuit width across most sizes, with tightly grouped medians that are generally lower than those of Qiskit. Qiskit tends to have a wider spread and higher median values, particularly for larger circuit sizes. The presence of more outliers in Qiskit’s data could suggest that it is either particularly sensitive to certain circuit configurations or it might exploit specific configurations more effectively than others.

To determine the algorithmic runtime, each QAOA circuit generated from the U3R graph configuration was evaluated seven times. The average runtime across evaluations was computed to provide a consistent measure of performance.

In evaluating the complexity of the three qubit reuse algorithms across various QAOA circuit sizes for \( p = 1 \), GidNET demonstrated significant performance superiority when compared to QNET and Qiskit as shown in ~\autoref{fig:qaoa_circuit_runtime_p1}. Our analysis showed that on average, GidNET operates approximately 87.9\% (or 8.3$\times$) faster than QNET and 98.1\% (or 52.6$\times$) faster than Qiskit. This substantial speed advantage highlights GidNET's ability in managing qubit resources, positioning it as a highly effective tool for qubit reuse.

~\autoref{fig:qaoa_circuit_width_p2_boxplot} compares the performance of GidNET and QNET qubit reuse algorithms on QAOA circuits with \( p=2 \). Both algorithms demonstrated similar effectiveness across circuit sizes, with GidNET showing a marginal advantage in achieving slightly smaller widths, especially as circuit complexity increased. Our result shows that on average, GidNET achieved  1.4\% improvement in circuit width reduction reaching up to 3.4\% for circuit size 42 over QNET. The variability in widths also grew with circuit size for both algorithms. We excluded the Qiskit qubit reuse algorithm from this analysis as its longer runtime made it impractical especially for larger circuits.

~\autoref{fig:qaoa_circuit_runtime_p2} compares the runtimes of GidNET and QNET
for QAOA circuits with \( p = 2 \). The analysis reveals that GidNET generally
demonstrates better runtime, particularly as the circuit size increases. GidNET
shows an average runtime improvement of 82.9\% (5.8$\times$) over
QNET. This performance becomes more pronounced for larger circuits: for 50-qubit circuits, we observe a runtime improvement of about 97.4\% (38.5$\times$). This pattern indicates that GidNET is particularly effective in managing the increased complexity of larger circuits, making it a potentially more suitable choice for extensive qubit reuse tasks where runtime is critical.

\subsection{Statistical Validation of Complexity Analysis}
Statistical analysis via polynomial regression was employed to evaluate how well the theoretically obtained complexity analyses of GidNET and QNET align with experimental results. This approach involved fitting polynomial models to the runtime data, with each model's coefficients tested to ascertain their impact. The primary objective was to validate the theoretical predictions against actual performance as circuit complexity increased. The selected degrees of the polynomials were based on reported computational complexities. QNET's complexity was previously established as \(O(mn + n^5)\) ~\cite{fang2023dynamic}, prompting our choice of a fifth-degree polynomial (\(n^5\)) for its analysis. Similarly, GidNET, with a reported complexity of \(O(mn^2 + \mu \nu k^2 + pn^3\log{n})\), was analyzed using a third-degree polynomial (\(n^3\)).

F-tests and R-squared values were employed to assess the models across various circuits. The results are summarized in ~\autoref{tab:statistical_results}. To avoid cluttering ~\autoref{fig:supremacy_runtime_comparison}, we show only the polynomial fit for depth 11 GRCS circuits for GidNET and QNET. The F-statistics were \(2874.84\) for GidNET and \(3725.56\) for QNET, indicating strong model fits, with R-squared values at \(0.9990\) and \(0.9996\) respectively, demonstrating that nearly all variance in the experimental data was accounted for. 

In the case of QAOA circuits with \(p=1\), GidNET's F-statistic was \(614.2\), and QNET's was \(19968.28\), with corresponding R-squared values of \(0.9999\) and \(0.99996\), indicating that both models provide a highly adequate fit to the reported complexities. For QAOA circuits with \(p=2\), GidNET was fitted with polynomial \(n^2\). This is because, although polynomial \(n^3\) and \(n^2\) have approximately the same R-squared value of \(0.998\),  \(n^3\) with F-statistic of \(800.95\) shows a worse fit compared to \(n^2\) with F-statistic of \(1117.60\). Also, depending on the circuit structure, it is common that not all the GidNET iterations are needed to achieve optimal reuse. QNET maintained a strong model fit with an F-statistic of \(1925.77\) and an R-squared value of \(0.9996\). These results not only substantiate the regression models' fit but also reinforce the credibility of the statistical validation by directly tying it to the theoretical framework of each algorithm.

\begin{table}[ht]
\centering
\caption{Statistical Analysis Results for GidNET and QNET on Various Circuits}
\label{tab:statistical_results}
\begin{tabular}{llcc}
\hline
Algorithm & Circuit & R-squared Value & F-statistic \\
\hline
GidNET & GRCS Circuit (\(d=11\)) & 0.9990 & 2874.84 \\
QNET & GRCS Circuit (\(d=11\)) & 0.9996 & 3725.56 \\
GidNET & QAOA Circuit (\(p=1\)) & 0.9971 & 614.2 \\
QNET & QAOA Circuit (\(p=1\)) & 1.0000 & 19968.28 \\
GidNET & QAOA Circuit (\(p=2\)) & 0.9976 & 1117.60 \\
QNET & QAOA Circuit (\(p=2\)) & 0.9996 & 1925.77 \\
\hline
\end{tabular}
\end{table}

We suspect the variations in the fit of polynomial models for QAOA and GRCS circuits can be attributed to the difference in circuit sizes used in the experiments. Specifically, QAOA circuits were tested with a smaller number of qubits compared to the GRCS circuits. The accuracy of the polynomial models tends to improve for circuits with more qubits.

\section{Conclusion and Future Work} \label{sec:conclusion}
This paper introduces GidNET (Graph-based Identification of qubit NETwork), a
novel algorithm aimed at optimizing qubit reuse in quantum circuits to manage
quantum resource constraints in current architectures. By leveraging the
circuit's DAG and candidate matrix, GidNET efficiently identifies pathways for
qubit reuse to reduce quantum resource requirements.

The core advantage of GidNET over other algorithms lies in its ability to offer
improved solutions with more scalable computational
demands. Our comparative analysis with existing algorithms, particularly QNET, shows that GidNET not only achieves smaller compiled circuit widths by an average of 4.4\%, with reductions reaching up to 21\% for larger circuits, but speeds up computations, achieving significant runtime improvements by an average of 97.4\% (or 38.5$\times$) reaching up to 99.3\% (or 142.9$\times$) across varying circuit sizes. This makes GidNET a highly effective choice for larger and more complex quantum circuits where both compilation time and qubit management are crucial factors. These results underscore the critical role of advanced qubit reuse strategies in optimizing quantum computations, providing a robust solution that bridges the gap between exact algorithmic precision and heuristic scalability, and a valuable benchmark for further development in this field.

GidNET, while offering  advancements in qubit reuse optimization for quantum circuits, is not without limitations. For example, the performance of GidNET could vary across different quantum hardware architectures, requiring specific tailoring to individual systems. Hence, future enhancements to GidNET could involve adapting it for specific quantum hardware architectures, and expanding its scope to multi-objective optimizations that consider not only circuit width but also depth and gate count. Additionally, integrating machine learning techniques could refine its graph-based strategies, potentially offering more dynamic and optimized qubit reuse sequences. A better method for updating the candidate matrix could result in improved complexity of GidNET making it possible to use GidNET for industry grade circuits requiring large number of qubits. These advancements could make GidNET a more versatile and powerful tool in optimizing quantum circuits for practical quantum computing applications.

\section*{Acknowledgments}
GU acknowledges funding from the NSERC CREATE in Quantum Computing Program, grant number 543245. TA acknowledges funding from NSERC. ODM is funded by NSERC, the Canada Research Chairs program, and UBC.

\bibliographystyle{IEEEtran}
\bibliography{references}

\end{document}